\documentclass[12pt,preprint]{aastex}
\newcommand{\bx}{{\bf x}}
\newcommand{\br}{{\bf r}}
\newcommand{\pr}{\prime}
\newcommand{\tC}{\tilde{C}}

\def\jcap{JCAP}
\def\beq{\begin{equation}}
\def\eeq{\end{equation}}
\def\ben{\begin{eqnarray}}
\def\een{\end{eqnarray}}
\def\numass{\sum m_{\nu}}
\def\lcdm{\Lambda{\rm CDM}}
\def\nucdm{\nu{\rm CDM}}
\def\frs{{\rm fR6}}
\def\frss{{\rm fR6+0.06}\,{\rm eV}}
\def\frf{{\rm fR5+0.15}\,{\rm eV}}
\def\sige{\sigma_{8}}
\def\ev{\,{\rm eV}}
\def\munit{\,h^{-1}\! M_{\odot}}
\def\dunit{\,h^{-1}{\rm Mpc}}
\def\rf{R_{f}}
\def\u{\hat{\bf u}}

\def\ej{\hat{\bf e}_{1}}
\def\ei{\hat{\bf e}_{2}}
\def\em{\hat{\bf e}_{3}}
\def\lamj{\hat{\bf \lambda}_{1}}
\def\lami{\hat{\bf \lambda}_{2}}
\def\lamm{\hat{\bf \lambda}_{3}}
\def\dt{d_{t}}
\def\etar{\eta (r)}
\def\cost{\vert\cos\theta\vert}
\usepackage{color}

\begin{document}
\title{Disentangling Modified Gravity and Massive Neutrinos with Intrinsic Shape Alignments of Massive Halos}
\author{Jounghun Lee\altaffilmark{1}, Suho Ryu\altaffilmark{1},  Marco Baldi\altaffilmark{2,3,4}}
\altaffiltext{1}{Astronomy Program, Department of Physics and Astronomy, FPRD, 
Seoul National University, Seoul 08826, Korea \email{shryu@astro.snu.ac.kr, jounghun@astro.snu.ac.kr}}
\altaffiltext{2}{Dipartimento di Fisica e Astronomia, Alma Mater Studiorum Universit\`a di Bologna, viale Berti Pichat, 
6/2, I-40127 Bologna, Italy}
\altaffiltext{3}{INAF - Osservatorio Astronomico di Bologna, via Ranzani 1, I-40127 Bologna, Italy}
\altaffiltext{4}{INFN - Sezione di Bologna, viale Berti Pichat 6/2, I-40127 Bologna, Italy}
\begin{abstract}
We present two new diagnostics based on the intrinsic shape alignments of group/cluster size dark matter halos to disentangle the effect of $f(R)$ gravity from that of massive neutrinos. 
Using the snapshot data from a series of the {\small DUSTGRAIN}-{pathfinder} $N$-body simulations for the Planck $\Lambda$CDM cosmology and 
three $f(R)$ gravity models with massive neutrinos ($\nu$), we first determine the probability density functions of the alignment angles between the shape orientations 
of massive halos and the minor principal axes of the local tidal fields.  The numerically obtained results turn out to agree very well with the analytic formula derived 
under the assumption that the anisotropic merging along the cosmic web induces the halo shape alignments. The four cosmologies, which several standard diagnostics 
failed to discriminate, are found to yield significantly different best-fit values of the single parameter that characterizes the analytic formula. 
We also numerically determine the spatial cross-correlations between the shape orientations of neighbor group/cluster halos, and find them to be in good agreements 
with a fitting formula characterized by two parameters, whose best-fit values are found to substantially differ among the four models. 
We also discuss the limitations and caveats of these new diagnostics that must be overcome for the application to real observational data. 
\end{abstract}
\section{Introduction}

If the scientific precept, occam's razor, were to be blindly used,  then the standard $\lcdm$ cosmology based on Einstein's general relativity (GR) with dominant cosmological 
constant ($\Lambda$) and cold dark matter (CDM) would win over all non-standard ones since it is the simplest and most effective paradigm within which almost all observations 
can be explained \citep[for a recent review, see][and references therein]{lcdm_ok}. 
Nevertheless, the notorious fine-tuning problem of $\Lambda$ \citep[see][for a review]{wei89,car01} has always left much room for us to come up with various alternatives in the faint 
hope that a more fundamental and natural description of the universe might become possible. 
Meanwhile, to survive multiple stringent observational tests, an alternative model must yield almost identical growth and expansion histories to the $\lcdm$ 
counterpart. 
In other words, the viability of an alternative cosmology is ensured only provided that it cannot be readily discriminated from the $\lcdm$ one by the conventional 
diagnostics based on the expansion and growth histories. 

For instance, some $f(R)$ gravity models combined with massive neutrinos ($\nu$) and CDM have been shown to be indistinguishable from the $\lcdm$ cosmology 
by such powerful standard probes as the density power spectrum, evolution of the cluster abundance, and matter-to-halo bias factor \citep{bal-etal14,hag-etal19a}. 
Here, $f(R)$ gravity is a modified gravity (MG) whose dynamics is described by the same Einstein-Hilbert action but with Ricci scalar, $R$, replaced by an arbitrary function, 
$f(R)$ \citep{buc70,sta80,HS07,LB07}. It predicts the presence of a fifth force in addition to gravity, but recovers GR in the high-density environments via a screening mechanism 
called the chameleon \citep{KW04}. The strength of a fifth force is quantified by the absolute value of its derivative, called the scalaron, at the  present epoch, 
$\vert f_{R0}\vert\equiv \vert df/dR\vert_{z=0}$ \citep[for a review, see][]{SF10}. 
The current observational constraint is $\vert f_{R0}\vert\lesssim 3.7\times10^{-6}$ on the cluster scale \citep[e.g.,][]{bou-etal14}. 

However, if the neutrinos ($\nu$) have non-zero total mass, $M_{\nu}>0$,  then the observational data could allow $\vert f_{R0}\vert$ to have larger values than this constraint, since the 
presence of massive neutrinos would simultaneously suppress structure formation, thereby counteracting the enhancement due to the fifth force. 
Some combination of $M_{\nu}$ with $\vert f_{R0}\vert$ can beguile the standard diagnostics not to detect the presence of free streaming neutrinos and fifth force by producing growth histories 
consistent with that of the standard $\lcdm$ case within observational uncertainties \citep{bal-etal14}. 
Breaking this {\it dark sector} degeneracy between the $\lcdm$ and the $\nucdm$+$f(R)$ cosmologies requires an acutely sensitive nonlinear diagnostics that can perceive and react to 
a subtle difference between them in spite of their very similar growth and expansion histories.  What has so far been proposed as such diagnostics includes the evolution of supercluster 
straightness, nonlinear growth rate, nonlinear redshift distortions, evolution of the drifting average coefficient of the isolated cluster abundance, high order weak lensing statistics, and 
high-redshift size distribution of cosmic voids \citep{pee-etal18,hag-etal19b,gio-etal19,wri-etal19,ryu-etal20,con-etal21}. 

Very recently, \citet{chu-etal21} suggested that the intrinsic shape alignment of galactic halos be useful as a probe of gravity, showing by $N$-body simulations that its strength 
significantly differ between the $\lcdm$ and a $f(R)$ gravity model with $f_{R0}=10^{-5}$ (fR5). Here, the intrinsic shape alignment of galactic halos refers to the phenomenon that 
the shape orientations of galactic halos exhibit a tendency of being preferentially aligned with the hosting filaments, which is believed to originate from the occurrence of anisotropic merging 
along the filamentary cosmic web \citep[][and references therein]{atl-etal06,hah-etal07,zha-etal09,zha-etal13,che-etal16,lee19}. 
The fR5 model that \citet{chu-etal21} mainly considered, however, has been already well known to be readily distinguishable from the $\lcdm$ cosmology by the 
aforementioned conventional diagnostics themselves \citep{bal-etal14}, since its growth and expansion histories significantly differ from that of the $\lcdm$ cosmology.  
In other words, it is not a surprise that the strength of the intrinsic shape alignments of galactic halos differs between the $\lcdm$ and the fR5 models. 
A critical question that arises in light of \citet{chu-etal21} is whether or not the halo shape alignment could provide a new independent clue to the nature of gravity and DM 
beyond the constraints put by those standard statistics.  

In this Paper, we intend to explore if the intrinsic shape alignments of group/cluster size halos can break the dark sector degeneracy between the $\lcdm$ and the $\nucdm$+$f(R)$ cosmologies. 
We choose the group/cluster size halos rather than the galactic counterparts, since the shape orientations of those massive halos were well known for long to exhibit much stronger intrinsic alignments 
with the filaments \citep[e.g.,][]{atl-etal06,hah-etal07,zha-etal13,lee19}. 
The plan of this paper is as follows. In Section \ref{sec:model}, we will briefly review the analytic prescriptions for the intrinsic shape alignments of DM halos with 
the tidal fields and present a new formula for the spatial correlations of the shape orientations of group/cluster halos. 
In Section \ref{sec:num}, we will numerically examine if and how efficiently two new diagnostics based on the intrinsic shape alignments of group/cluster size halos 
can break the dark sector degeneracy among the $\lcdm$ and three different $\nucdm$+$f(R)$ cosmologies. 
In Section \ref{sec:con}, we will summarize the results and discuss the advantages and disadvantages of these new diagnostics compared with the conventional ones. 

\section{A Brief Review of the Analytic Model}\label{sec:model}

Suppose that a DM halo is found at some position, ${\bf x}$, where the local tidal tensor, ${\bf T}=(T_{ij})$, smoothed on the scale of $\rf$, has three eigenvectors, 
$\{{\bf e}_{1},{\bf e}_{2},{\bf e}_{3}\}$, corresponding to three eigenvalues, $\{\lambda_{1},\lambda_{2},\lambda_{3}\vert \lambda_{1}\ge\lambda_{2}\ge\lambda_{3}\}$. 
The eigenvectors, ${\bf e}_{1}$ and ${\bf e}_{3}$, are parallel to the directions of maximum and minimum matter compression, respectively. 
Since the sum of the three eigenvalues of ${\bf T}$ is equal to the local density, $\delta$, the traceless tidal tensor, $\tilde{\bf T}=(\tilde{T}_{ij})$, can be obtained as 
$\tilde{T}_{ij}\equiv T_{ij}- (\delta/3)I_{ij}$, where $I_{ij}$ is the $3\times 3$ identity matrix. To single out only the anisotropic effect of the tidal field on the orientations of DM halo shapes, 
we will deal with $\tilde{\bf T}$ rather than ${\bf T}$. 

Suppose also that the DM halo has an ellipsoidal shape whose inertia momentum tensor, ${\bf U}=(U_{ij})$, has three eigenvalues and corresponding eigenvectors.
Let $\u$ denote a unit vector parallel to the major eigenvector of $(U_{ij})$ corresponding to its largest eigenvalue. We will refer to $\u$ as the halo shape orientation 
throughout this paper.  If a halo formed through an anisotropic merging along the cosmic web, its shape orientation, $\u$, is expected to be in the direction of 
minimum matter compression.  
To quantitatively describe this expected alignment of $\u$ with ${\bf e}_{3}$, \citet{lee19} proposed the following analytic formula for the probability density 
function of $\cost\equiv\vert\u\cdot\em\vert$:
\ben
\label{eqn:cost}
p(\cost) &=& \frac{1}{2\pi}\int_{0}^{2\pi}\left[\prod_{n=1}^{3}\left(1+d_{t}-3d_{t}\hat{\lambda}_{n}\right)\right]^{-\frac{1}{2}}
\left(\sum_{l=1}^{3}\frac{\vert\hat{\bf u}\cdot\hat{\bf e}_{l}\vert}{1+d_{t}-3d_{t}\hat{\lambda}_{l}} \right)^{-\frac{3}{2}}\,d\phi\, .
\een
Here $\{\hat{\bf e}_{1},\hat{\bf e}_{2},\hat{\bf e}_{3}\}$ are the orthonormal eigenvectors of the unit traceless tidal tensor, $\hat{\bf T}\equiv \tilde{\bf T}/\vert{\tilde T}\vert$, 
in parallel to $\{{\bf e}_{1},{\bf e}_{2},{\bf e}_{3}\}$ corresponding to its three eigenvalues 
$\{\hat{\lambda}_{1},\hat{\lambda}_{2},\hat{\lambda}_{3}\vert \hat{\lambda}^{2}_{1}+\hat{\lambda}^{2}_{2}+\hat{\lambda}^{2}_{3}=1\}$, 
$\theta$ and $\phi$ are the spherical polar and azimuthal angles of $\hat{\bf u}$, respectively, in the principal frame of $\hat{\bf T}$, 
and $\dt$ is an empirical parameter introduced to measure the strength of the $\u$-$\em$ alignment tendency.  
\citet{lee19} showed that the best-fit value of $\dt$ varies with $R_{f}$, reaching its maximum possible value when 
$R_{f}\equiv\left[3M/(4\pi\rho_{m})\right]^{1/3}$, where $\rho_{m}$ is the mean density of the universe, and $M$ is the halo mass. This optimal smoothing scale 
amounts to four times the halo virial radius, $R_{f}\sim 4r_{\rm vir}$ \citep{lib-etal13a}.

If the shape orientations of DM halos are truly aligned with the direction of minimum matter compression parallel to $\em$, then it is expected that the large-scale coherence of 
$\hat{\bf T}$ would induce the shape-shape alignments between the neighbor halos.  \citet{lee-etal09} defined the spatial cross-correlation function of $\u$ as 
\begin{equation}
\eta(r) = \langle\vert\u(\bx)\cdot\u({\bx}+{\br})\vert^{2}\rangle - \frac{1}{3}\, ,
\label{eqn:eta}
\end{equation}
where $\br$ is a separation vector, $1/3$ is the value of $\langle\vert\u(\bx)\cdot\u({\bx}+{\br})\vert^{2}\rangle$ in the asymptotic limit that $r\equiv {\br}/\vert{\br}\vert$ 
goes to infinity.  The ensemble average in Equation (\ref{eqn:eta}) is to be taken over the halo pairs separated by the same distance, $r$.

The functional behavior of $\eta (r)$ should be closely linked with the spatial cross-correlations of the tidal fields, 
$C_{ijkl}(\br)\equiv \langle T_{ij}(\bx) T_{kl}(\bx + \br)\rangle$,  
the analytic expression for which was derived by \citet{LP01} in the linear limit where ${\bf T}$ can be described as a Gaussian random field:
\ben  
C_{ijkl}(\br) &=&(I_{ij}I_{kl}+I_{ik}I_{jl}+I_{il}I_{jk})\bigg{\{}\frac{J_3}{6} - \frac{J_5}{10}\bigg{\}} + 
(\hat{r}_i\hat{r}_j\hat{r}_k\hat{r}_l)\bigg{\{}\xi(r) + \frac{5J_3}{2} - \frac{7J_5}{2}\bigg{\}} + \nonumber \\
&&\frac{1}{2}(I_{ij}\hat{r}_k\hat{r}_l + I_{ik}\hat{r}_j\hat{r}_l
+ I_{il}\hat{r}_k\hat{r}_j + I_{jk}\hat{r}_i\hat{r}_l
+ I_{jl}\hat{r}_i\hat{r}_k + I_{kl}\hat{r}_i\hat{r}_j )\bigg{\{}J_5-J_3\bigg{\}}\, .
\label{eqn:tcor}
\een
Here $\xi (r)$ is the two-point correlation function of the linear density field, and $J_3 (r)$ and $J_5 (r)$ represent the third and fifth moments of $\xi (r)$, respectively, 
defined as  \citep{LP01}, 
\ben 
\label{eqn:xi}
\xi (r) &=& \frac{1}{2\pi^{2}}\int_{0}^{\infty} P(k) W^{2}(k, \rf) \frac{\sin kr}{kr}k^{2}dk\, ,  \\
\label{eqn:j3}
J_3 (r) &\equiv& \frac{3}{r^{3}}\int_0^r \xi(r^{\pr}) r^{\pr 2} dr^{\pr 2}\, , \\
\label{eqn:j5}
J_5 (r) &\equiv& \frac{5}{r^{5}}\int_0^r \xi(r^{\pr}) r^{\pr 4} dr^{\pr 4}\, , 
\een
where $P(k)$ is the linear density power specturm, and $W(k, \rf)$ is the top-hat window function with a filtering radius of $\rf$. 
The spatial cross-correlations of the traceless tidal tensor, 
$\tilde{C}_{ijkl}(\br)\equiv \langle\tilde{T}_{ij}(\bx)\tilde{T}_{kl}(\bx + \br)\rangle$, can be obtained from Equation (\ref{eqn:tcor}) as 
\ben
\tC_{ijkl}(\br) = C_{ijkl} - \frac{1}{3}I_{kl}C_{ijnn} - \frac{1}{3}I_{ij}C_{mmkl} + \frac{1}{9}I_{ij}I_{kl}C_{mmnn}\, .
\label{eqn:tlcor}
\een
Equation (\ref{eqn:tlcor}) reveals that the spatial correlation of $\tilde{\bf T}$ in the linear regime depends on the background cosmology through two linear quantities, 
$\xi (r)$ and $J_{3} (r)-J_{5} (r)$.  If $\xi (r)$ behaved like a power-law function of $r$, then $\tilde{C}_{ijkl}$ would depend only on $\xi (r)$ since $J_{3}$ would be equal to $J_{5}$. 
The large-scale coherence of the anisotropic tidal field, which deviates $\xi (r)$ from the simple power-law scaling, plays the role of linking $\tilde{C}_{ijkl}$ to $J_{3} (r)-J_{5} (r)$.  

Proposing an ansatz that the spatial cross-correlations of the halo shape orientations can be expressed in terms of $\xi (r)$ and $J_{3} (r)-J_{5} (r)$ like $\tilde{C}_{ijkl}$,  
we put forth the following two-parameter fitting formula for $\eta (r)$:
\begin{equation}
\eta(r) = g_{1}\frac{\xi (r;\rf)}{\sigma^{2}(\rf)} + g_{2}\left[\frac{J_{3}(r;\rf)}{\sigma^{2}(\rf)}-\frac{J_{5}(r;\rf)}{\sigma^{2}(\rf)}\right]\, ,
\label{eqn:eta_fit}
\end{equation}
where $\sigma (\rf)$ is the rms fluctuation of the linear density field smoothed on the scale of $\rf$,  $g_{1}$ and $g_{2}$ are two free parameters introduced to measure the strength 
of the cross-correlations of the halo shape orientations. The first parameter, $g_{1}$, measures the strength of the shape-shape alignments between the halos with comparable masses, 
while the second one, $g_{2}$, between the halos with different masses. Like $\dt$ in Equation (\ref{eqn:cost}), these two parameters are expected to vary with the smoothing scale, 
having their maximum possible values on the scale of $R_{f}\sim 4r_{\rm vir}$ \citep{lib-etal13a}. 

\section{Numerical Analysis}\label{sec:num}

We utilize a data subset from the {\small DUSTGRAIN}-{\it pathfinder} project \citep{gio-etal19}, a series of DM only $N$-body experiments that simulated various $\nucdm$+$f(R)$ 
cosmologies as well as a $\lcdm$ cosmology, with the help of the {\small MG-GADGET} code \citep{mggadget}. 
For the {\small DUSTGRAIN}-{\it pathfinder} simulations, the widely applied Hu-Sawicki formula was used to specify $f(R)$ \citep{HS07}, and the scheme developed by 
\citet{vie-etal10} was applied to treat the massive neutrinos as hot DM particles.  The linear box size ($L_{\rm box}$), the total number of DM particles ($N_{\rm par}$) 
and the particle mass resolution ($m_{\rm par}$) of the simulations are $L_{\rm box}=750\dunit$, $N_{\rm par}=768^{3}$, and $m_{\rm par}=8.1\times 10^{10}\munit$, respectively. 
We refer the readers to \citet{gio-etal19} and \citet{mggadget} for full descriptions of the {\small DUSTGRAIN}-{\it pathfinder} project and the {\small MG-GADGET} code, 
respectively. 

For our analysis, we consider the Planck $\lcdm$ \citep{planck16} and three different $\nucdm$+$f(R)$ cosmologies, namely, the $\frs$ with $\vert f_{R0}\vert=10^{-6}$ 
and $M_{\nu}=0.0\ev$, the $\frss$ with $\vert f_{R0}\vert=10^{-6}$ and $M_{\nu}=0.06\ev$, and the $\frf$ with $\vert f_{R0}\vert=10^{-5}$ and $M_{\nu}=0.15\ev$. 
These three $\nucdm$+$f(R)$ models have been known to be highly degenerate with the Planck $\lcdm$ cosmology by the standard diagnostics based on the growth history 
\citep{bal-etal14,hag-etal19a}. Especially, the $\frss$ and $\lcdm$ pair shares the identical normalization amplitude of the linear density power spectrum, $\sige$, 
while the $\frf$ and $\frs$ pair exhibits negligibly small difference in $\sige$ between each other (see Table \ref{tab:sig}). 

For each of the four cosmologies, we apply the Rockstar algorithm \citep{rockstar} to the DM particle snapshot at $z=0$ to obtain a catalog of DM halos which lists  
a variety of their properties such as its position (${\bf x}$), virial mass ($M$) and shapes. Basically, the method devised by \citet{all-etal06} was incorporated into the 
Rockstar algorithm to determine the ellipsoidal shape of a DM halo, providing information on the direction of the shape orientation of each halo, $\u$, as well as on the 
intermediate to major axial ratio, $p$. 
Applying the mass-cut $M_{\rm cut}=8.1\times 10^{12}\munit$ and axial ratio cut $p_{\rm cut}=0.9$ to the virial mass and intermediate to major axial ratios, 
we make a selection of massive triaxial distinct halos without being embedded in any larger halos. These two conditions are to ensure relatively high degree of 
accuracy in the determination of the shape orientations of the selected halos\footnote{As shown in \citet{all-etal06},  the minimum number of particles, $1000$, 
corresponding to the mass cut $M_{\rm cut}=8.1\times 10^{13}\munit$ for the current analysis, is in fact required to ensure negligible degree of numerical contamination 
in the determination of the halo shapes. However, the current analysis cannot afford to this high value of $M_{\rm cut}$ since the number of the halos would drastically drop  
due to the relatively low particle resolution of the simulations.}. 
The fifth column of Table \ref{tab:sig} also shows the number of the selected massive halos, $N_{8.1}$, for each model. 
Employing the ${\bf T}$-reconstruction routine described in \citet{lee19}, we determine the unit traceless tidal tensor, 
$\hat{\bf T}$, smoothed on the scale of $\rf$ at the position of each selected halo. Via the similarity transformation, we find three eigenvalues, $\lamj$, $\lami$, $\lamm$, 
and corresponding eigenvectors, $\ej$, $\ei$, and $\em$ of $\hat{\bf T}$ \citep[see also][]{lee-etal21}. 

For each selected halo, we calculate $\cost\equiv\vert\u\cdot\em\vert$. Splitting the range of $0\le\cost\le1$ into twelve short bins of equal length, $\Delta \cos\theta$, and 
counting the numbers of those halos whose values of $\cost$ fall in each bin, $\Delta N$, we numerically determine the probability density function, 
$p(\cost)=\Delta N/(\Delta\cos\theta\times N_{\rm tot})$ where $N_{\rm tot}$ denotes the total number of the selected massive halos at $z=0$. 
Plugging the mean values of $\lamj$, $\lami$, $\lamm$ averaged over the selected halos into Equation (\ref{eqn:cost}), we fit the analytic formula to the numerical results with 
Poissonian errors by adjusting the parameter, $\dt$, via the $\chi^{2}(d_{t})$-minimization. The error in $d_{t}$ is calculated as one standard deviation, $\sigma_{d_t}$, 
from the maximum likelihood distribution, $p(d_{t})\propto\exp\left[-\chi(d_{t})^{2}/2\right]$. 

\begin{figure}[ht]
\begin{center}
\plotone{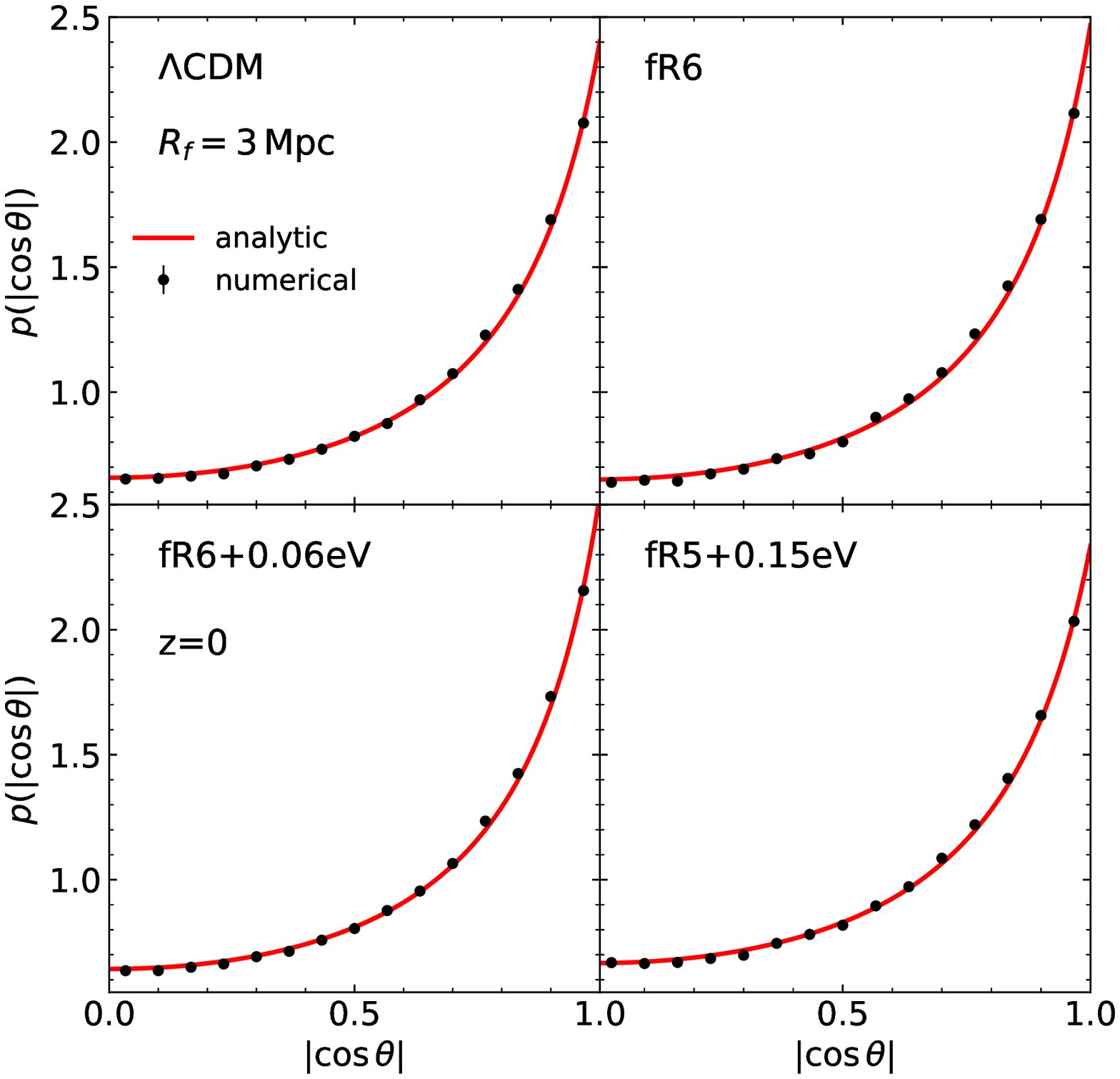}
\caption{Numerically obtained probability density functions of the absolute values of the cosines of the angles between the cluster 
shapes and the directions of minimum matter compression (black filled circles with errors) compared with the analytic model with 
one best-fit parameter (red solid line), on the scale of $\rf=3\dunit$ at $z=0$ for four different cosmologies.}
\label{fig:align_rf3}
\end{center}
\end{figure}

\begin{figure}[ht]
\begin{center}
\plotone{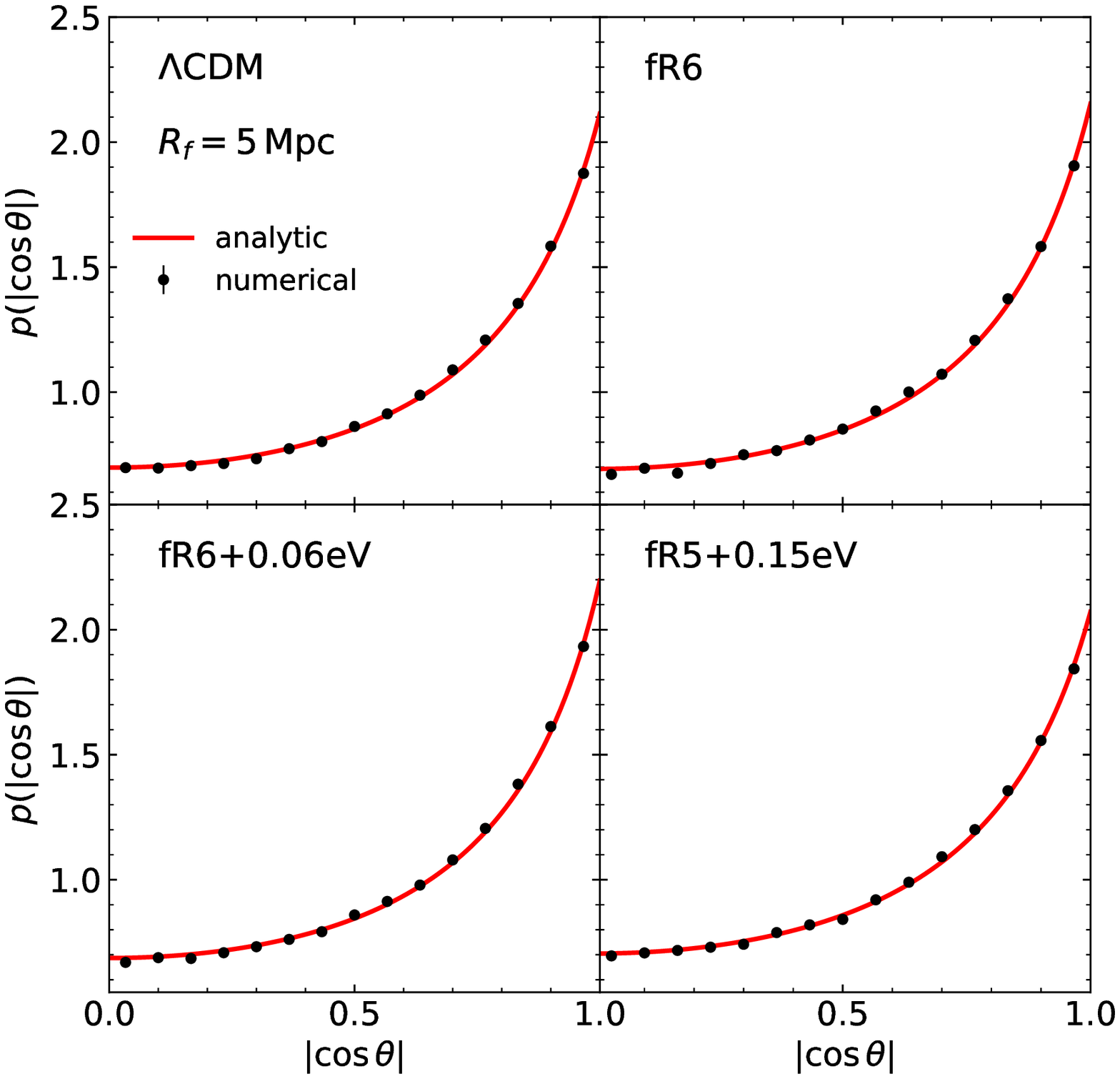}
\caption{Same as Figure \ref{fig:align_rf3} but on the scale of $\rf=5\dunit$.}
\label{fig:align_rf5}
\end{center}
\end{figure}

\begin{figure}[ht]
\begin{center}
\plotone{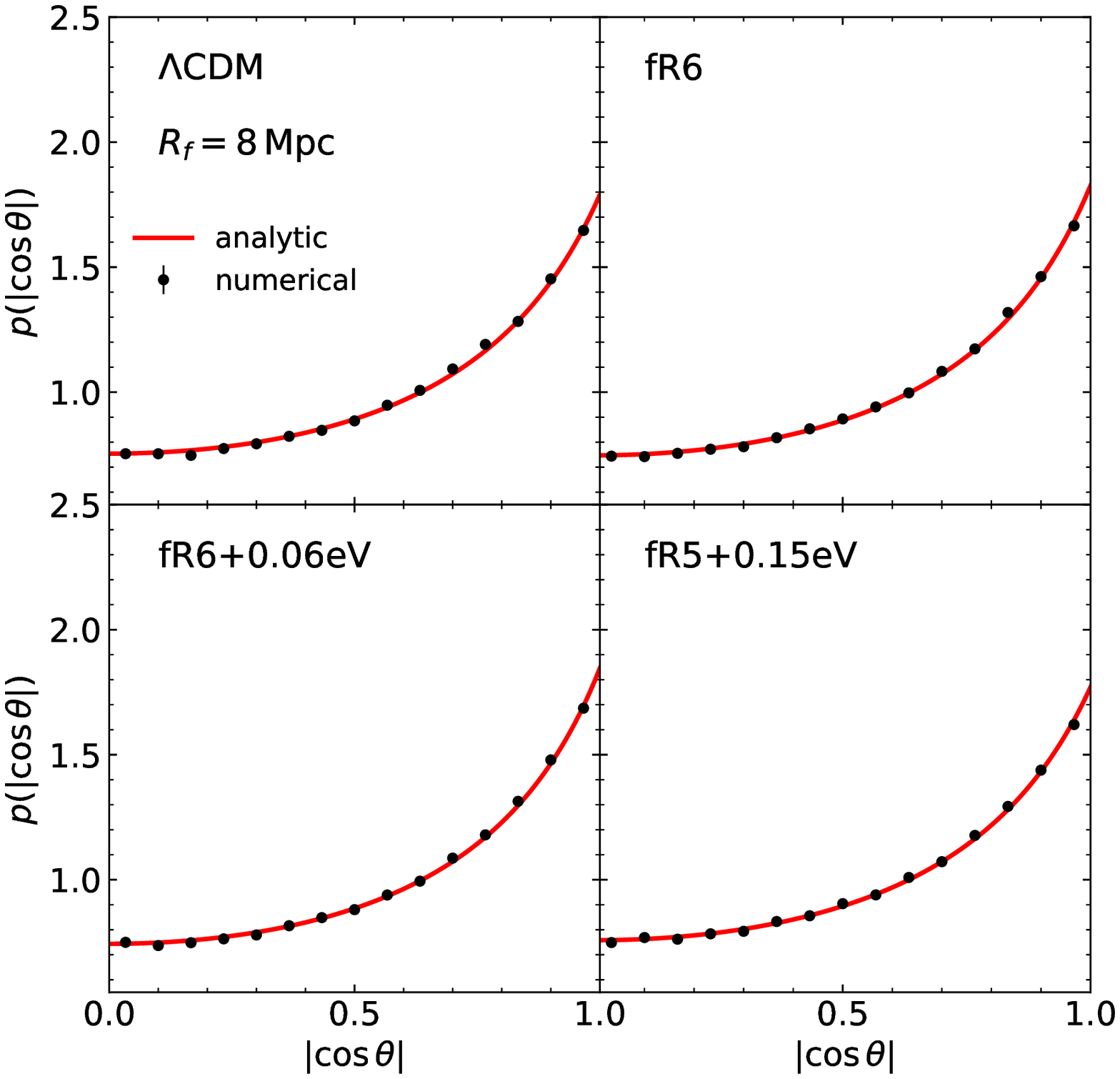}
\caption{Same as Figure \ref{fig:align_rf3} but on the scale of $\rf=8\dunit$.}
\label{fig:align_rf8}
\end{center}
\end{figure}

Figure \ref{fig:align_rf3} plots the numerically obtained $p(\cost)$ (black filled circles) with Poisson errors as well as the analytic formula, Equation (\ref{eqn:cost}), with 
the best-fit parameter (red solid line) on the scale of $\rf=3\dunit$ at $z=0$ for the four different cosmological models. This choice of $\rf=3\dunit$ is made by 
compromising between the particle resolution of the simulations and the validity of Equation (\ref{eqn:cost}). As can be seen, all of the four models exhibit the 
existence of strong $\u$-$\em$ alignments, but differ from one another in the functional behavior of $p(\cost)$, which is excellently described by the analytic formula with 
the best-fit value of $d_{t}$. We also repeat the same calculation but for two different cases of the smoothing scales,  $\rf=5\dunit$ and $8\dunit$, the results of which 
are shown Figures \ref{fig:align_rf5}-\ref{fig:align_rf8}, respectively. The good agreements between the numerical and analytical results seem to be robust against the variation 
of $\rf$. For a given cosmology, however, the $\u$-$\em$ alignment becomes stronger for the case that the tidal field is smoothed on the smaller scale, in consistent 
with what \citet{lee19} found for the $\lcdm$ case. 

\begin{figure}[ht]
\begin{center}
\plotone{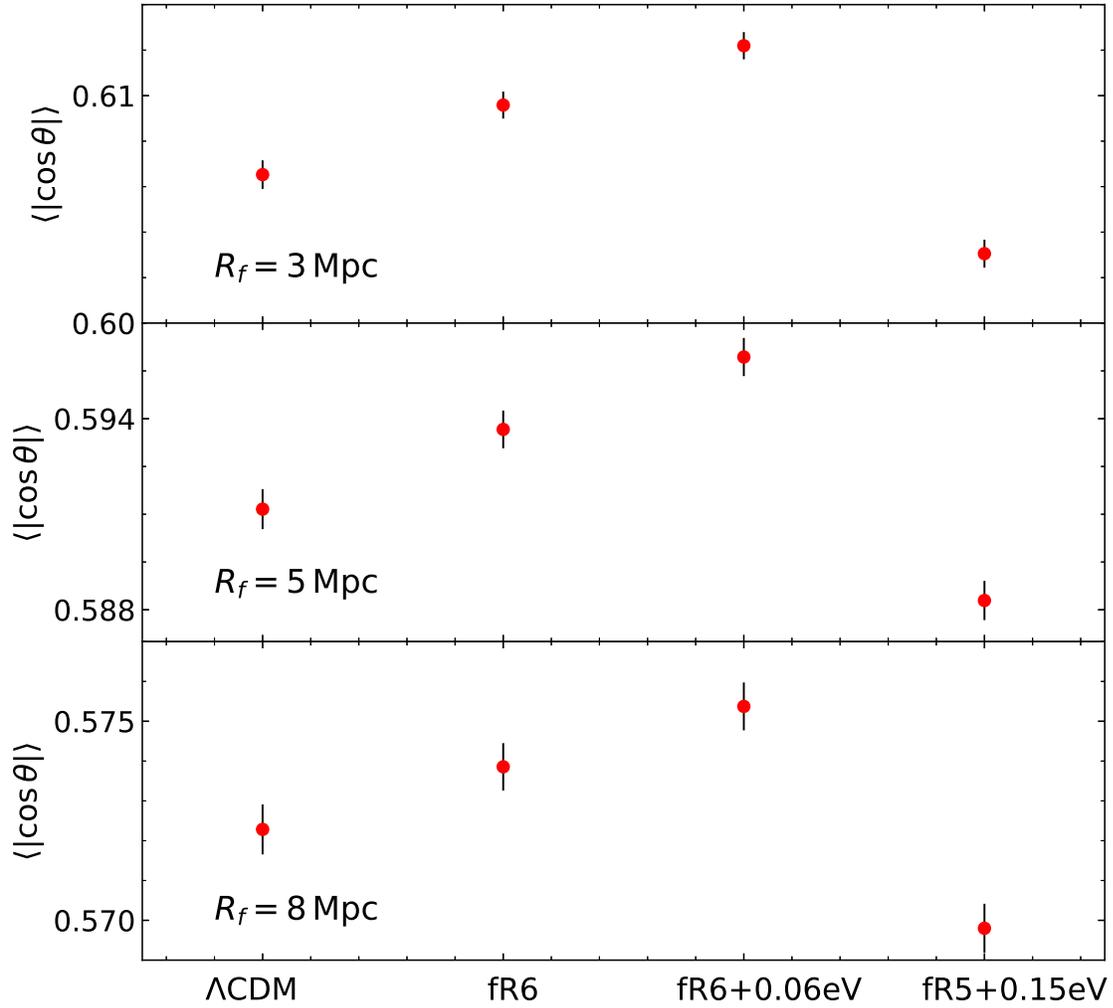}
\caption{Numerically obtained mean absolute values of the cosines of the angles between the cluster shapes 
and the directions of minimum matter compression on three different scales for four different cosmologies.}
\label{fig:malign}
\end{center}
\end{figure}

\begin{figure}[ht]
\begin{center}
\plotone{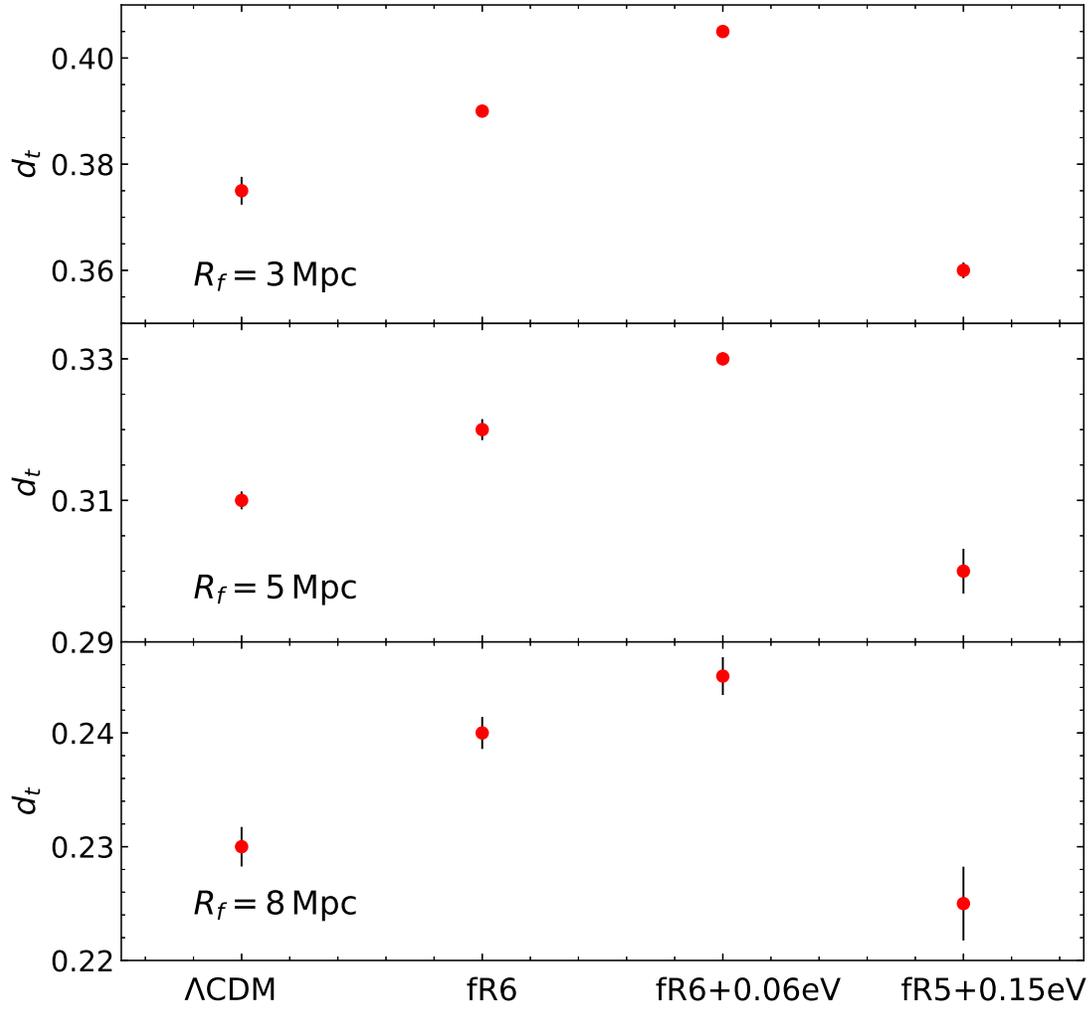}
\caption{Best-fit values of the shape correlation parameter on three different scales for four different cosmologies.}
\label{fig:dt}
\end{center}
\end{figure}

Figures \ref{fig:malign} and \ref{fig:dt} show how significantly the four models differ from one another in the ensemble average, $\langle\cost\rangle$, and in the best-fit value 
of $\dt$, respectively, for the three cases of $\rf$. Note that the single parameter, $d_{t}$, of the analytic formula varies more strongly with the four models than 
$\langle\cost\rangle$, demonstrating the efficiency of the analytic formula in distinguishing among the four models. 
The success of Equation (\ref{eqn:cost}) in matching the numerical results and its efficiency in discriminating the degenerate models implies that the nature of gravity and 
DM must leave a unique imprint on the shape orientations of group/cluster halos relative to the large-scale tidal fields by modulating the anisotropic occurrence of 
the merger events along the cosmic web. 

Our explanation for the non-monotonic or incoherent change of $\dt$ with $\vert f_{R}\vert$ and $\numass$, is as follow. The strength of the shape 
alignments of group/cluster halos with the tidal field is determined not only by how rapidly the density fluctuations grow but also by how anisotropic 
the large-scale cosmic web becomes in the nonlinear regime. The incoherent change of $\dt$ simply reflects the fact that the strength of this 
alignment is the consequence of the delicate counter-balance between the two effects. The high value of $\vert f_{R}\vert$ enhances the density growth, spurring the hierarchical 
merging, which can enhance the shape alignments.  But, at the same time, the high value of $\vert f_{R}\vert$ makes the cosmic web less anisotropic 
(i.e., more isotropic), which can weaken the shape alignments. Similarly, the high value of $\numass$ can have similar dual effects on the shape alignments since the 
presence of more massive neutrinos can suppress more severely the density growth, deterring the hierarchical merging process, while it can also make the cosmic web 
more anisotropic. 
 
The monotonic increase of $\dt$ as the background changes from the $\lcdm$ to the $\frss$ model implies that in these ranges of $0\le \vert f_{R}\vert\le 10^{-6}$ 
and $0\le \numass/{\rm eV}\le 0.06$, the gravity and neutrinos are strong and massive enough, respectively, to have the enhancing effect on the shape alignments. 
The lowest value of $\dt$ exhibited by the $\frf$ model implies that at these high values of $\vert f_{R}\vert=10^{-5}$ and $\numass/{\rm eV}=0.15$ the gravity and neutrinos 
are so strong and so massive, respectively,  that they have the opposite effect of reducing the shape alignments.  This incoherent variation of $\dt$ with $\vert f_{R}\vert$ and 
$\numass$ is in fact a manifestation of its potential to discriminate the degenerate cosmological models, which both of the linear and nonlinear density 
correlations fail to achieve. In other words, our statistics can complement the other conventional diagnostics based on the density growth history since it bring out 
a different aspect of the nonlinear evolution of the halos in the cosmic web: depending not only on how fast the halos grow but also on how anisotropic merging process 
the halos undergo. 

Although it turns out that $p(\cost)$ is very powerful in principle to distinguish among the degenerate models,  the practical success of this 
statistics is contingent upon the availability of prior information on the background cosmology for the reconstruction of the real-space tidal fields.
Another diagnostics based on the shape orientations of DM halos, which does no require such prior information, is their spatial cross-correlations, $\eta(r)$. 
For the numerical determination of $\eta(r)$, we calculate the separation distance, $r$, between each pair of the selected halos with shape orientations, say, $\u_{1}$ and $\u_{2}$.  
Dividing the range of $r$ into multiple bins of equal length $\Delta r = 2\,h^{-1}$Mpc, we compute the ensemble average, $\langle\vert\u_{1}\cdot\u_{2}\vert^{2}- 1/3\rangle$ 
over the halo pairs whose separation distances fall in $[r,\ r+\Delta r]$ to obtain $\etar$. Then, we create $1000$ bootstrap resamples of the halo pairs falling in each of the 
$r$ bins, and calculate one standard deviation scatter among the resamples as the associated error, $\sigma_{\eta}$. 

The analytic formula for $\eta (r)$, Equation (\ref{eqn:eta_fit}), is fitted to the numerically obtained $\eta(r)$ by adjusting the two parameters, $g_{1}$ and $g_{2}$, 
via the $\chi^{2}(g_{1},g_{2})$-minimization.  To evaluate $\xi (r)$, $J_{3}(r)$ and $J_{5}(r)$ in Equation (\ref{eqn:eta_fit}), we consistently use the same linear density 
power spectrum of the Planck $\lcdm$ model for all of the four models, mimicking the realistic situation where no prior information on the background cosmology is available. 
If $\eta (r)$ is truly a powerful discriminator of non-standard models, then the best-fit values of $g_{1}$ and $g_{2}$ would significantly differ among the four models, 
despite that the same linear density power spectrum is consistently used to evaluate the analytic formula. 
 
\begin{figure}[ht]
\begin{center}
\plotone{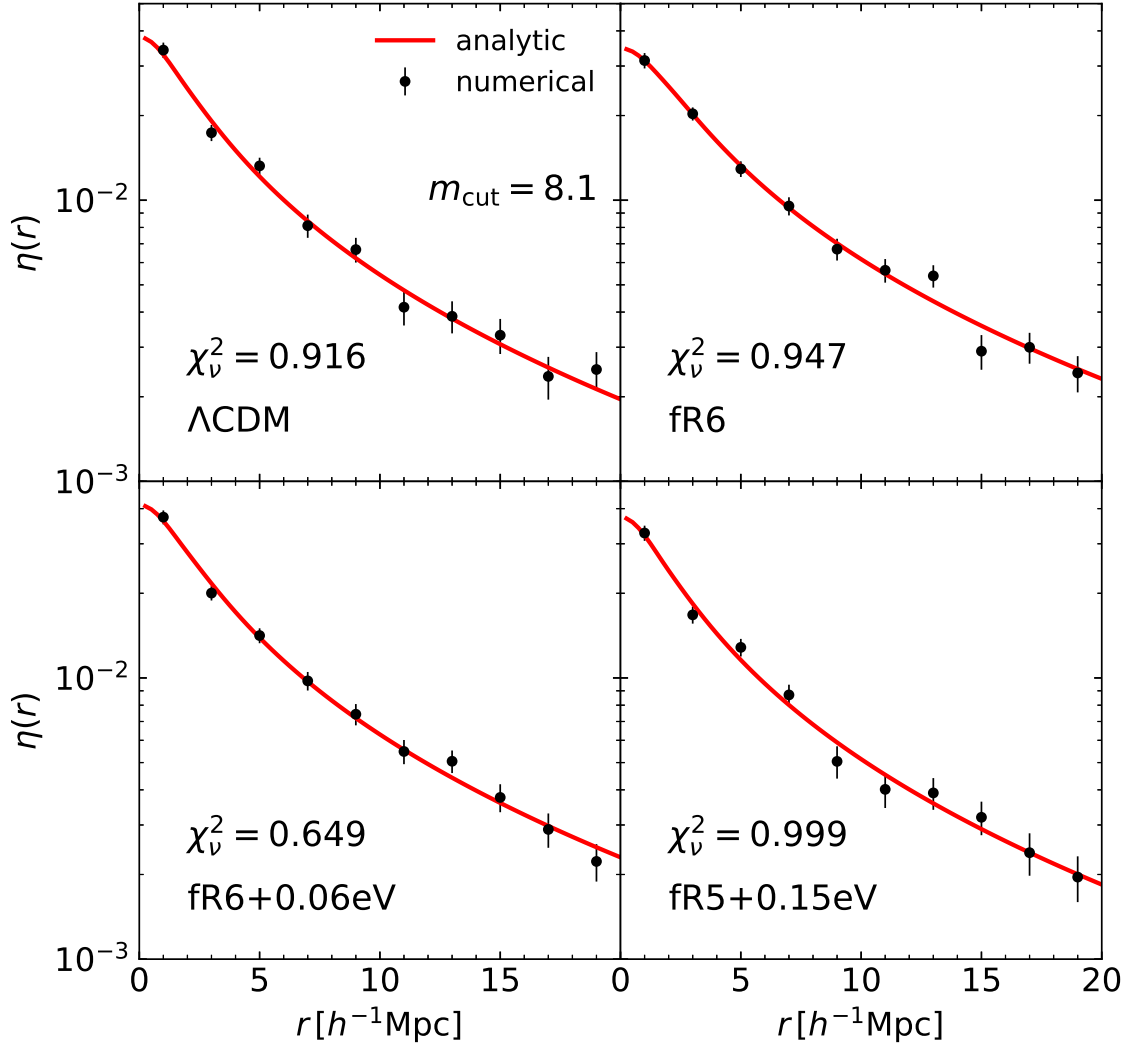}
\caption{Numerically obtained halo shape-shape cross-correlation functions (black filled circles) compared with 
the analytic model with two best-fit parameters (red solid curves) for four different cosmologies at $z=0$.}
\label{fig:cross}
\end{center}
\end{figure}

\begin{figure}[ht]
\begin{center}
\plotone{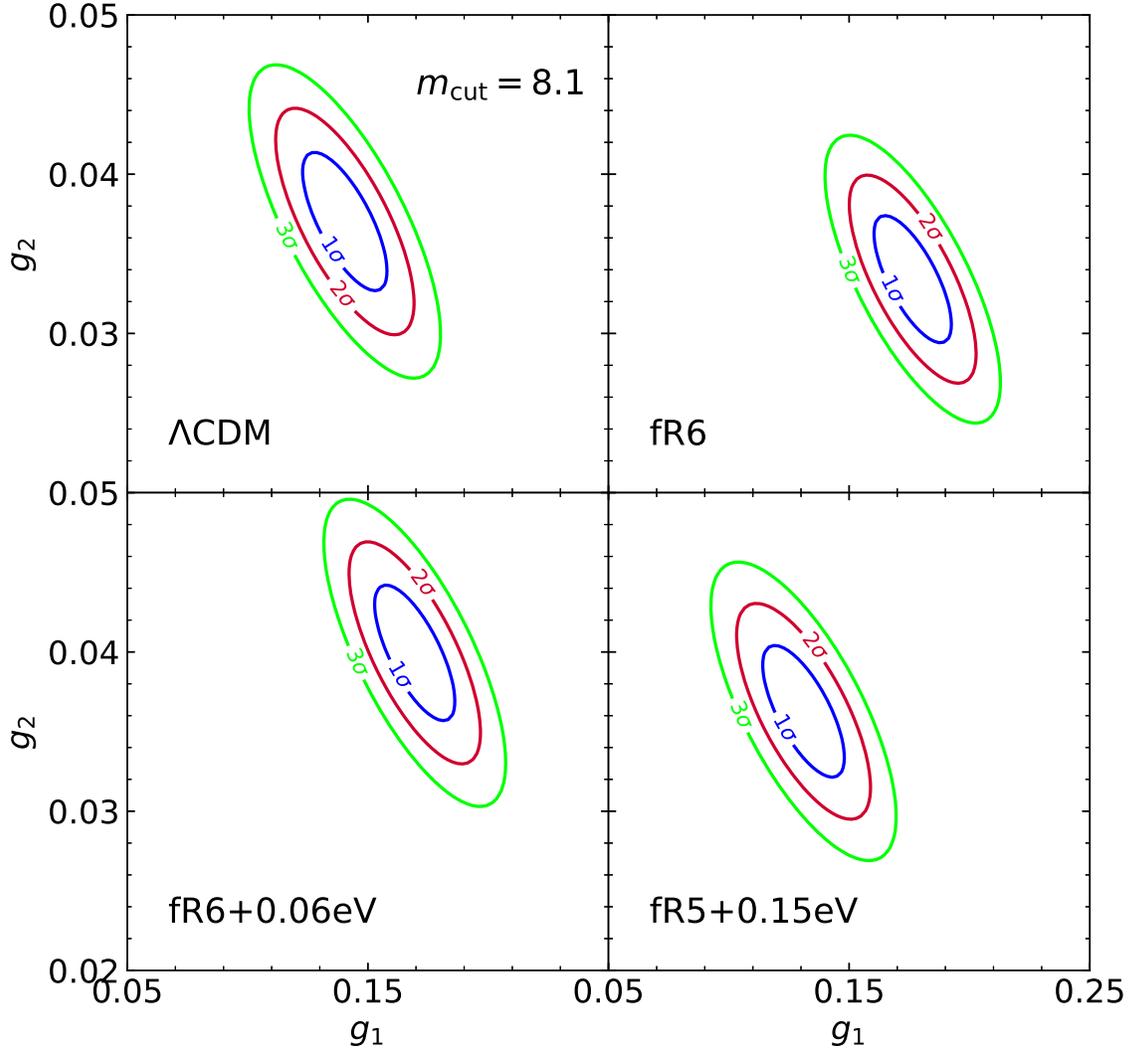}
\caption{$68\%$, $95\%$ and $99\%$ confidence regions of the $\chi^{2}$-values in the configuration space 
spanned by the two shape correlation parameters.}
\label{fig:cont}
\end{center}
\end{figure}

Figure \ref{fig:cross} plots the numerically derived $\etar$ (black filled circles) at $z=0$ as well as the analytic formula (red solid line) with the best-fit parameters, 
$g_{1}$ and $g_{2}$, for the four models. It also shows the values of  the reduced $\chi^{2}$, i.e., $\chi^{2}_{\nu}\equiv\chi^{2}/n_{f}$, where $n_{f}$ denotes the degree of 
freedom equal to the number of bins subtracted by $2$, the number of the free parameters.  As can be seen, for each model, we detect a clear signal of strong spatial cross-correlation 
of the halo shape orientations over a large distance up to $20\dunit$, and find good agreements between the numerical results and the analytic formula with the best-fit parameters for all 
of the four models. Figure \ref{fig:cont} plots the $68\%$, $95\%$ and $99\%$ contours of $\chi^{2}(g_{1},g_{2})$ for the four models. 

\begin{figure}[ht]
\begin{center}
\plotone{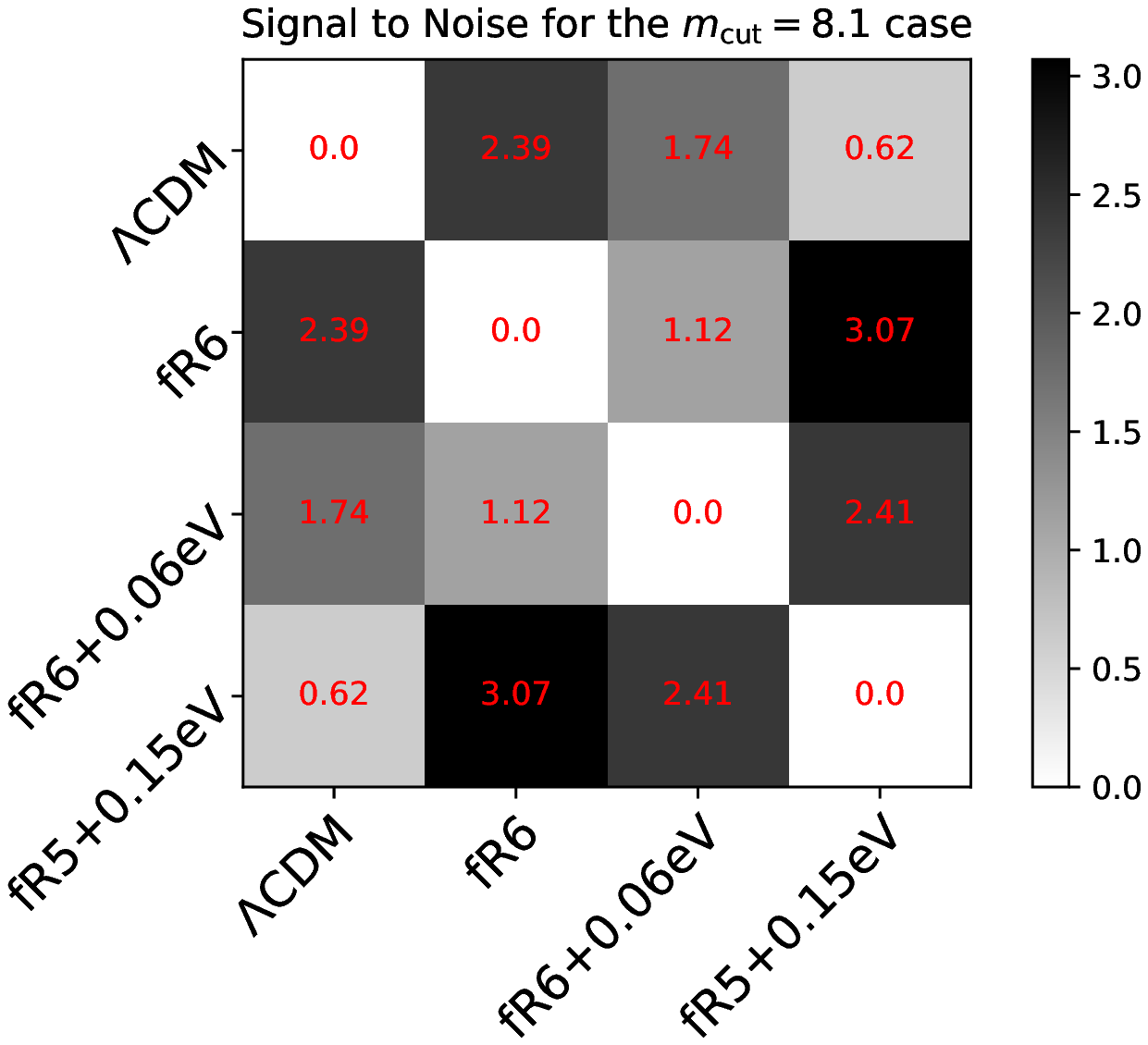}
\caption{Signal to noise ratios of the differences among the $\lcdm$, $\frs$, $\frss$, and $\frf$ cosmologies.}
\label{fig:cov}
\end{center}
\end{figure}

The statistical differences in $\eta(r)$ between the two degenerate models can be effectively quantified by measuring the distances in the 
configuration space spanned by the two parameters. Let $\{g_{1},\ g_{2},\ \sigma_{g_{1}},\  \sigma_{g_{2}},\sigma_{g_{1}g_{2}}\}$ and 
$\{g^{\prime}_{1},\ g^{\prime}_{2},\ \sigma_{g^{\prime}_{1}},\  \sigma_{g^{\prime}_{2}}, \sigma_{g^{\prime}_{1}g^{\prime}_{2}}\}$ denote two sets of the first and second best-fit parameters 
of $\eta$, their marginalized errors and the covariance between them for the two models, respectively, which can be all obtained from the numerically obtained posterior distributions 
$p[-\chi^{2}(g_{1},g_{2})/2]$. Defining the distance $D$ between them as $D\equiv \left[(g_{1}-g^{\prime}_{1})^{2}+(g_{2}-g^{\prime}_{2})^{2}\right]^{1/2}$, we determine the associated 
errors by the following {\it error propagation formula} as
\begin{eqnarray}
\label{eqn:sigd}
\sigma^{2}_{D}&=&\left(\frac{\partial D}{\partial g_{1}}\right)^{2}\sigma^{2}_{g_{1}} + \left(\frac{\partial D}{\partial g_{2}}\right)^{2}\sigma^{2}_{g_{2}} 
+ \left(\frac{\partial D}{\partial g^{\prime}_{1}}\right)^{2}\sigma^{2}_{g^{\prime}_{1}} + \left(\frac{\partial D}{\partial g^{\prime}_{2}}\right)^{2}\sigma^{2}_{g^{\prime}_{2}}\, 
+ \nonumber \\
&& 2\left(\frac{\partial D}{\partial g_{1}}\right)\left(\frac{\partial D}{\partial g_{2}}\right)\sigma_{g_{1}g_{2}} + 
2\left(\frac{\partial D}{\partial g^{\prime}_{1}}\right)\left(\frac{\partial D}{\partial g^{\prime}_{2}}\right)\sigma_{g^{\prime}_{1}g^{\prime}_{2}}\, . 
\end{eqnarray}
Figure \ref{fig:cov} plots $D/\sigma_{D}$ for each pair of the four cosmologies.  The two degenerate models, $\frs$ and $\frf$, yield the highest signal to noise ratio, 
$D/\sigma_{D}\gtrsim 3$,  while the other two degenerate models, $\lcdm$ and $\frss$, produce only $1.74$ signal-to-noise ratio. Note also that the substantial signal to noise ratios are 
found from the other two pairs:  $D/\sigma_{D}=2.39$ ($D/\sigma_{D}=2.41$) from the $\lcdm$ and $\frs$ ($\frss$ and $\frf$) models. 

\begin{figure}[ht]
\begin{center}
\plotone{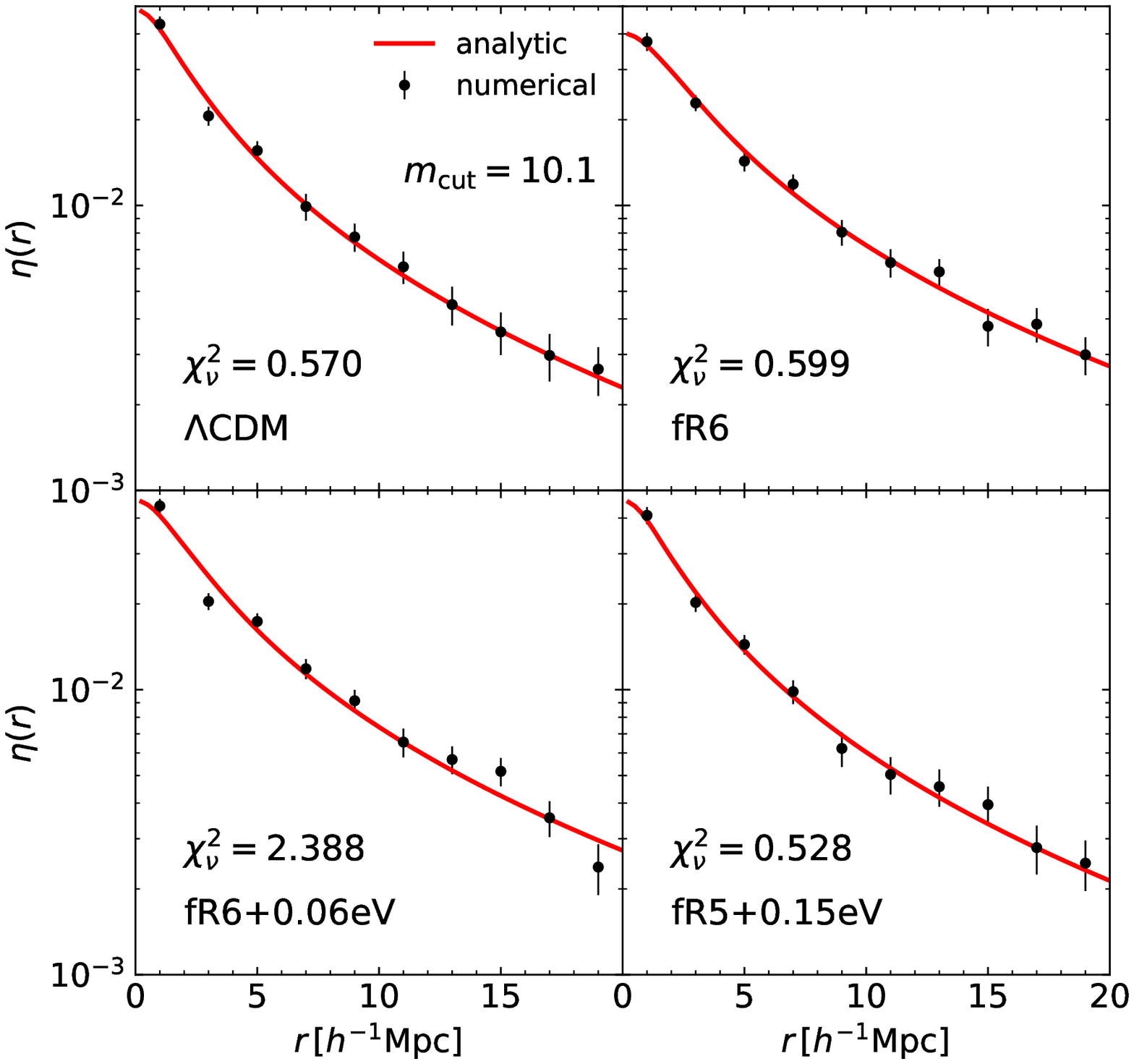}
\caption{Same as Figure \ref{fig:cross} but for the case of a higher mass-cut, $10.1\times 10^{12}\munit$, is applied.}
\label{fig:cross10}
\end{center}
\end{figure}

\begin{figure}[ht]
\begin{center}
\plotone{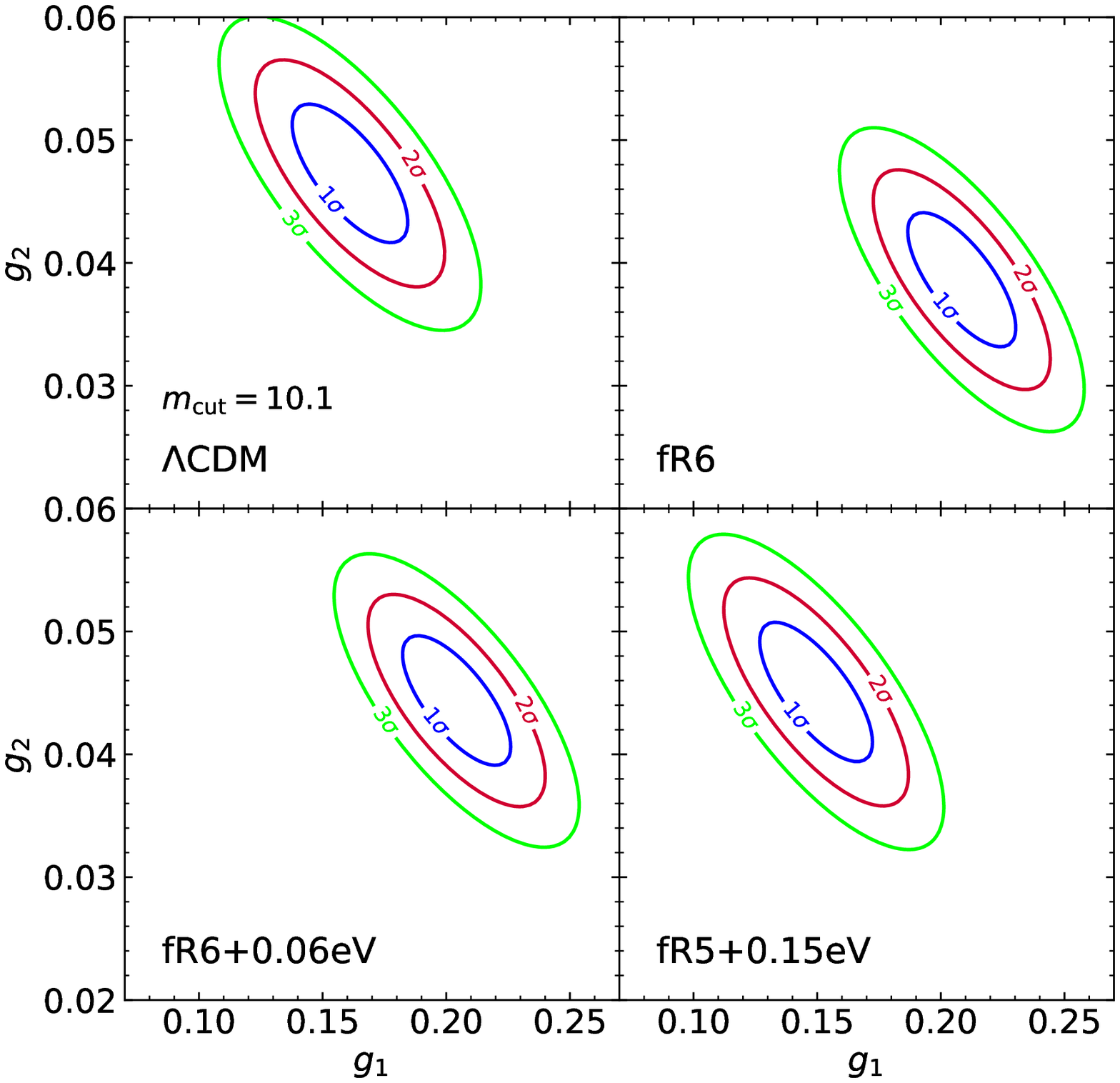}
\caption{Same as Figure \ref{fig:cont} but for the case of a higher mass-cut, $10.1\times 10^{12}\munit$, is applied.}
\label{fig:cont10}
\end{center}
\end{figure}

\begin{figure}[ht]
\begin{center}
\plotone{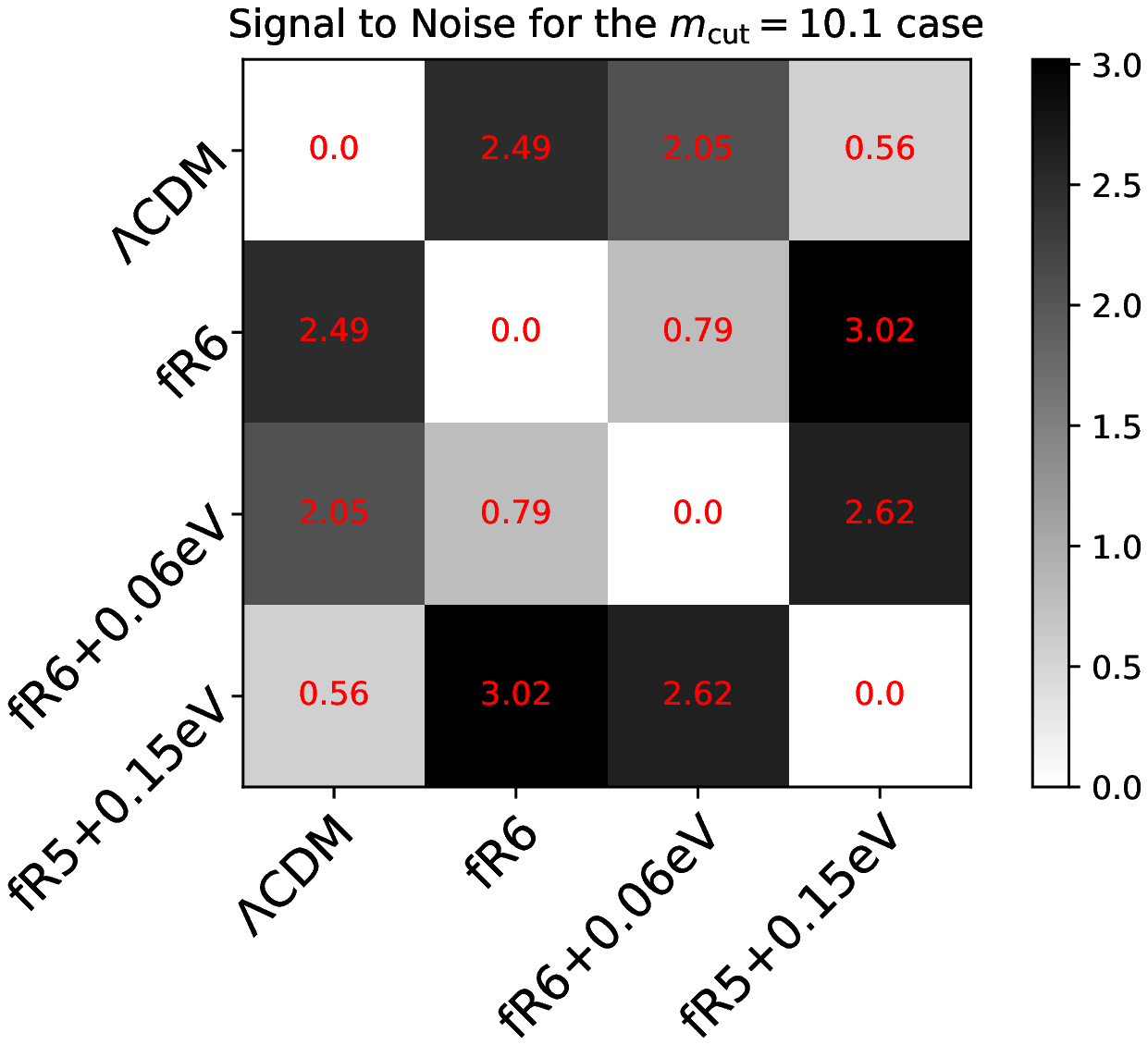}
\caption{Same as Figure \ref{fig:cov} but for the case of a higher mass-cut, $10.1\times 10^{12}\munit$, is applied.}
\label{fig:cov10}
\end{center}
\end{figure}

To see if the result would depend on our specific choice of $M_{\rm cut}$ for the selection of the halos, we repeat the whole analysis but with a higher mass-cut, 
$M_{\rm cut}=10.1\times 10^{12}\munit$.  The number of the halos, $N_{10.1}$, selected by applying this higher mass-cut for each model is  provided in the sixth column of Table \ref{tab:sig}. 
Figures \ref{fig:cross10}-\ref{fig:cov10} plot the same as Figures  \ref{fig:cross}-\ref{fig:cov} but from the halos selected by applying $M_{\rm cut}=10.1\times 10^{12}\munit$, 
respectively. As can be seen from Figure \ref{fig:cross10}, the analytic formula, Equation (\ref{eqn:eta_fit}), still works quite 
well in describing the numerical results, demonstrating its robustness against the variation of $M_{\rm cut}$. The comparison ot Figure \ref{fig:cont} with Figure \ref{fig:cont10} 
also proves the consistency in the trends of the two parameters with the four models except for the contour sizes in spite of the change of $M_{\rm cut}$. 
Figure \ref{fig:cov10} reveals that this new statistics based on $\eta$  is in principle capable of breaking the degeneracy between the 
$\frs$ and $\frf$ models, still producing significant signal to noise ratio, $D/\sigma_{D}>3$. 

\begin{figure}[ht]
\begin{center}
\plotone{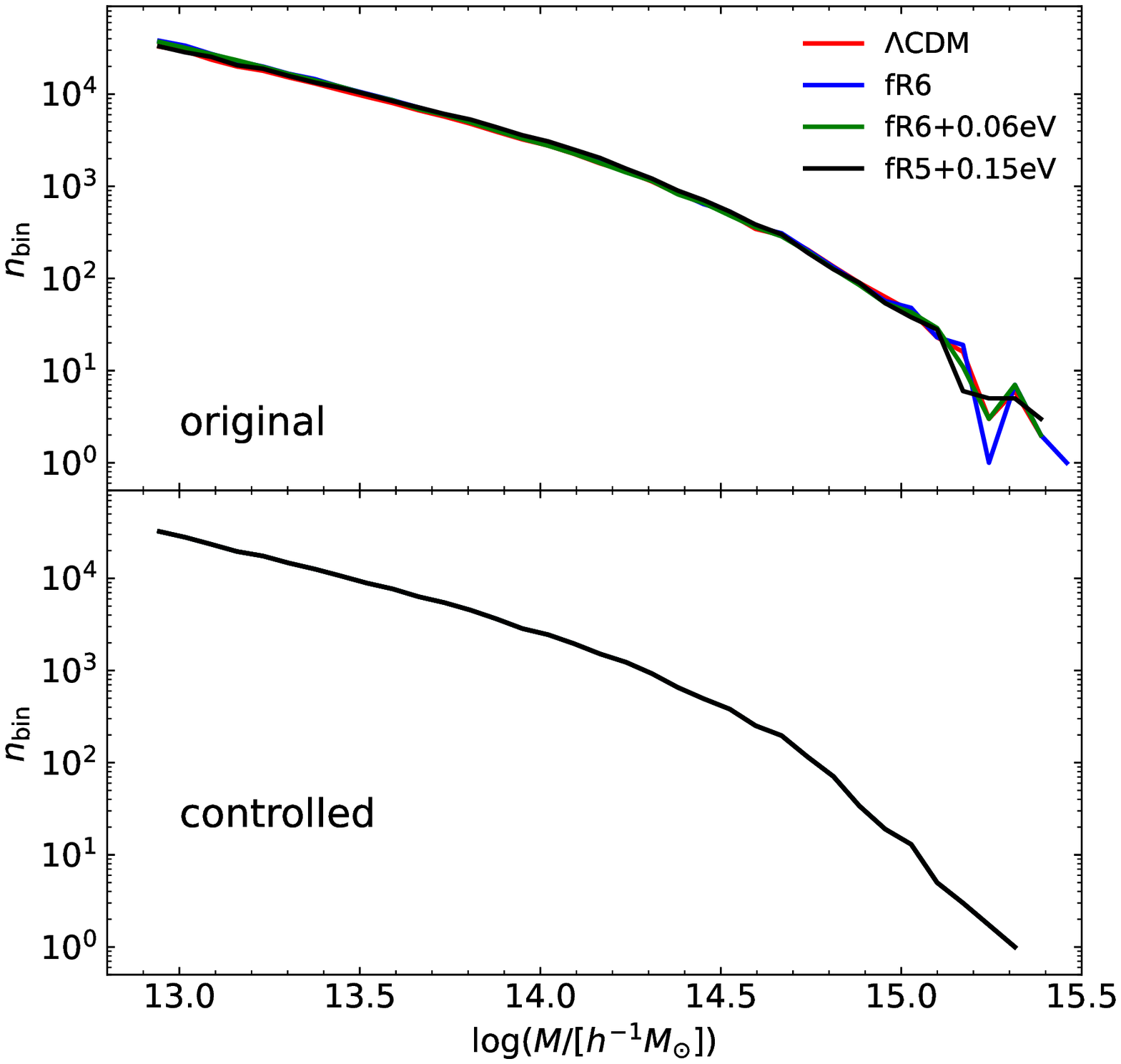}
\caption{Number distributions of the halos belonging the original (top panel) and controlled (bottom panel) samples 
from the four models as a function of the halo mass.}
\label{fig:mdis}
\end{center}
\end{figure}

\begin{figure}[ht]
\begin{center}
\plotone{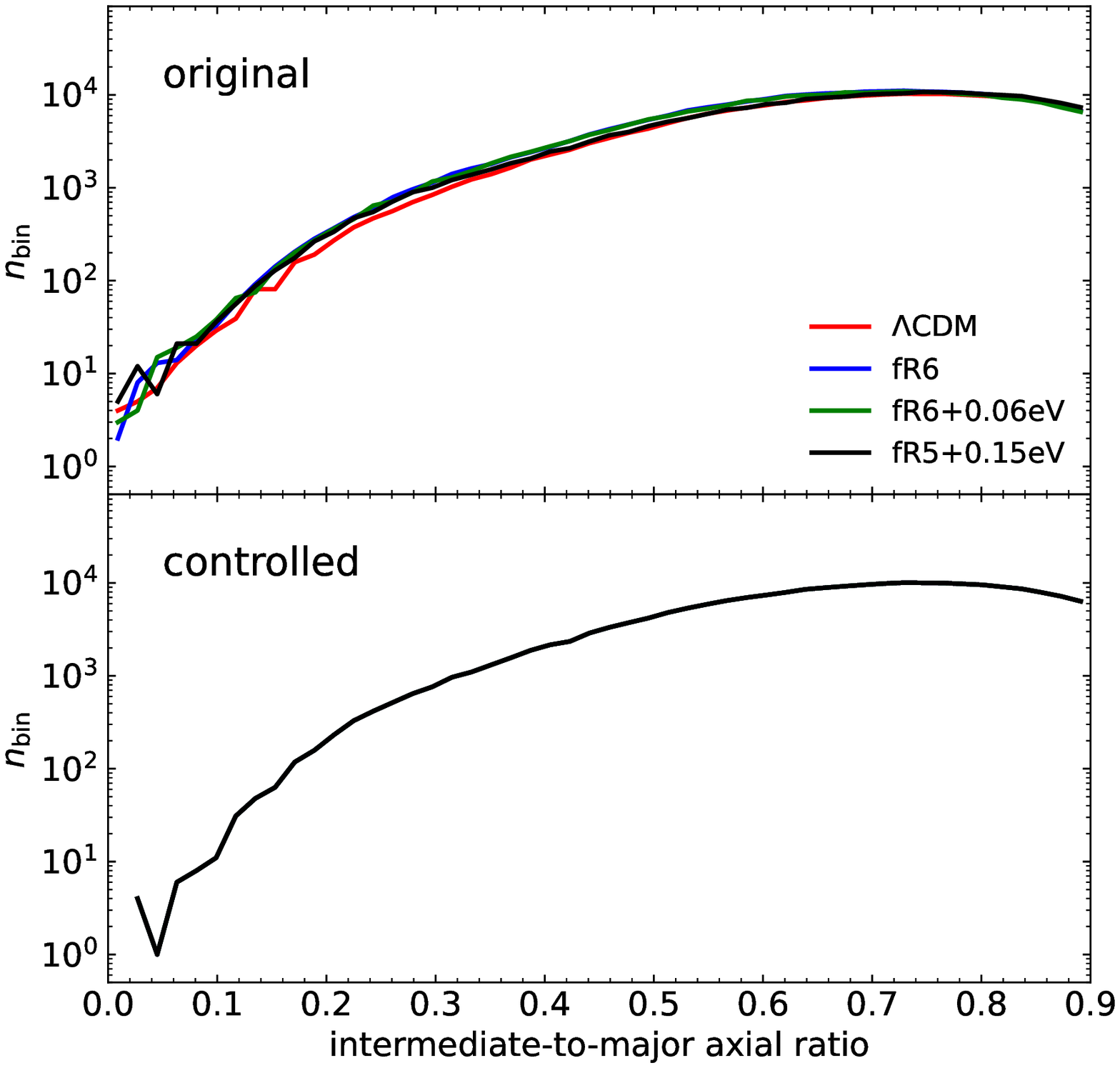}
\caption{Number distributions of the halos belonging the original (top panel) and controlled (bottom panel) samples 
from the four models as a function of the halo axial ratios.}
\label{fig:rdis}
\end{center}
\end{figure}

There is, however, one more test that $\eta (r)$ must pass for its validity to be confirmed as a discriminator of the degenerate cosmologies. 
Even though the mass and axial ratio distributions of the four degenerate models are quite similar to one another, they are not identical.  Since the shape-shape correlations 
are strongly dependent on the halo mass and axial ratios,  it should be necessary to inspect whether or not the detected difference in $\eta(r)$ among the degenerate models are 
due to the difference in the mass and axial ratio distributions. For this inspection, we control the halo samples from the four models to share the identical mass 
and axial ratio distributions. The seventh column of Table \ref{tab:sig} also lists the total number of the halos, $N_{\rm sync}$, belonging to the controlled samples for each model. 
Figures \ref{fig:mdis}-\ref{fig:rdis} plot the number counts of the halos as a function of their mass and axial ratios, respectively from the original (top panel) and controlled 
(bottom panel) subsamples. As can be seen, the small but non-negligible differences in the mass and axial ratio distributions among the four models disappear as the controlled 
subsamples replace the original ones. 

\begin{figure}[ht]
\begin{center}
\plotone{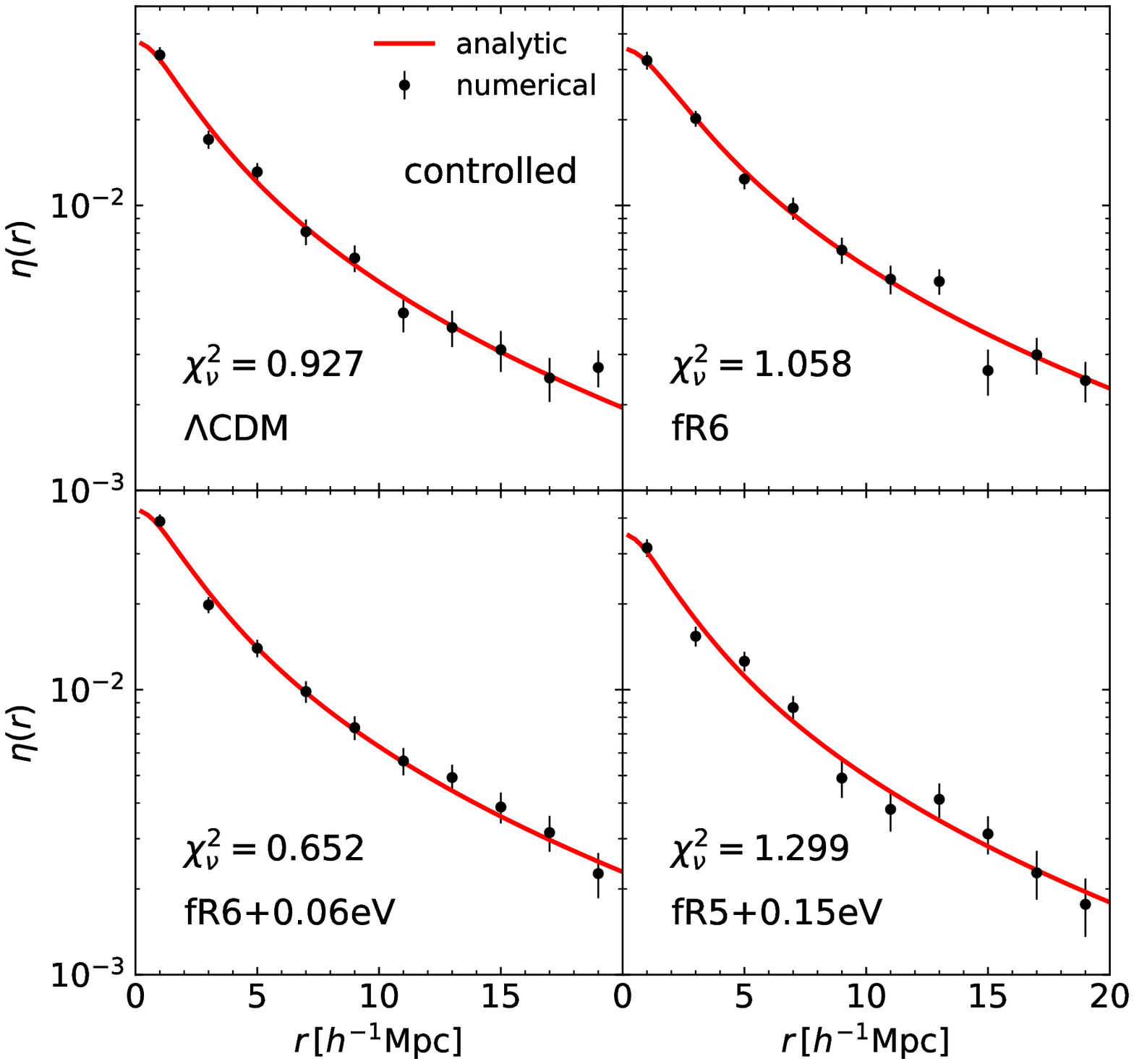}
\caption{Same as Figure \ref{fig:cross} but from the controlled samples.}
\label{fig:cross_sync}
\end{center}
\end{figure}

\begin{figure}[ht]
\begin{center}
\plotone{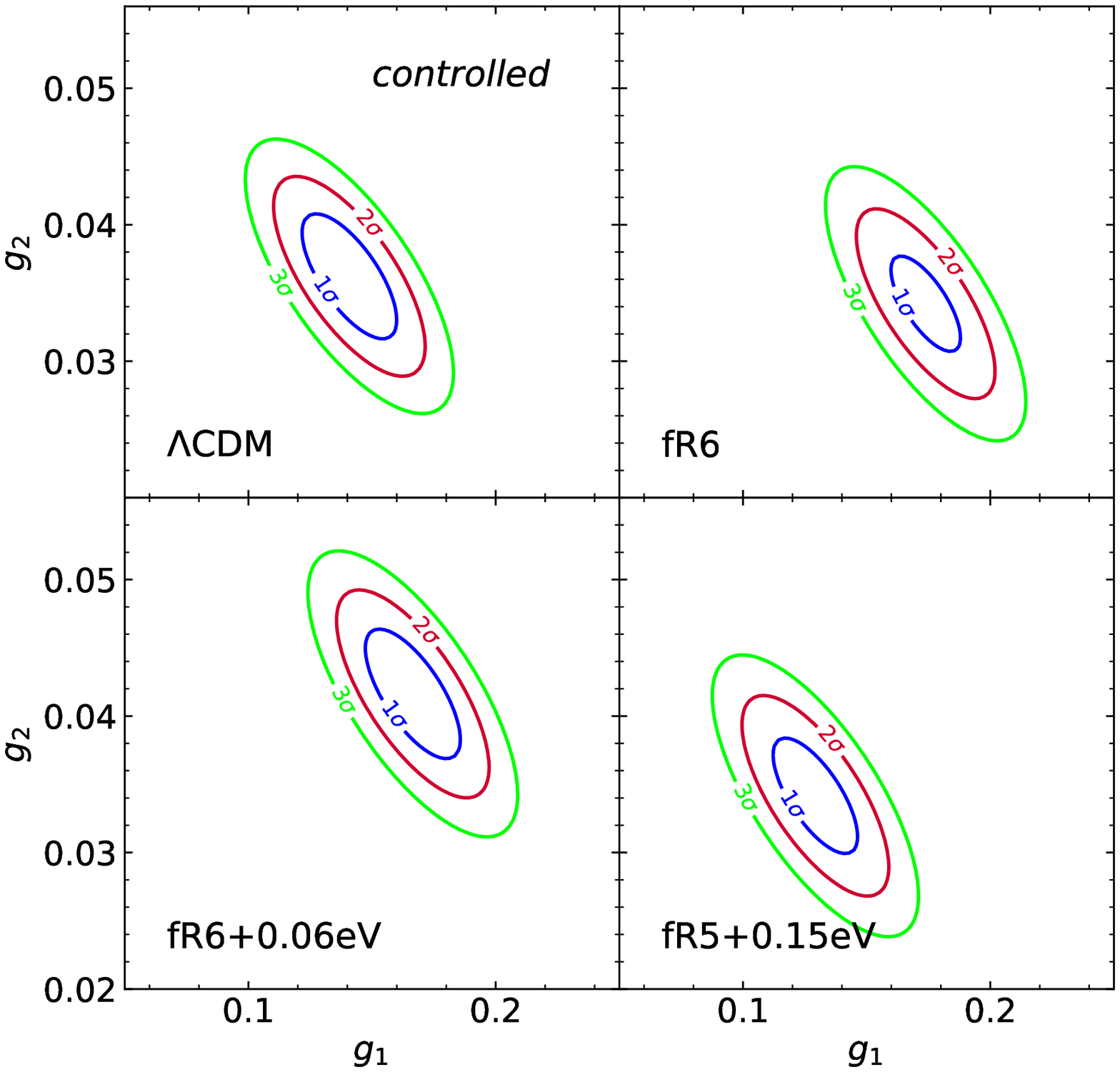}
\caption{Same as Figure \ref{fig:cont} but from the controlled samples.}
\label{fig:cont_sync}
\end{center}
\end{figure}

\begin{figure}[ht]
\begin{center}
\plotone{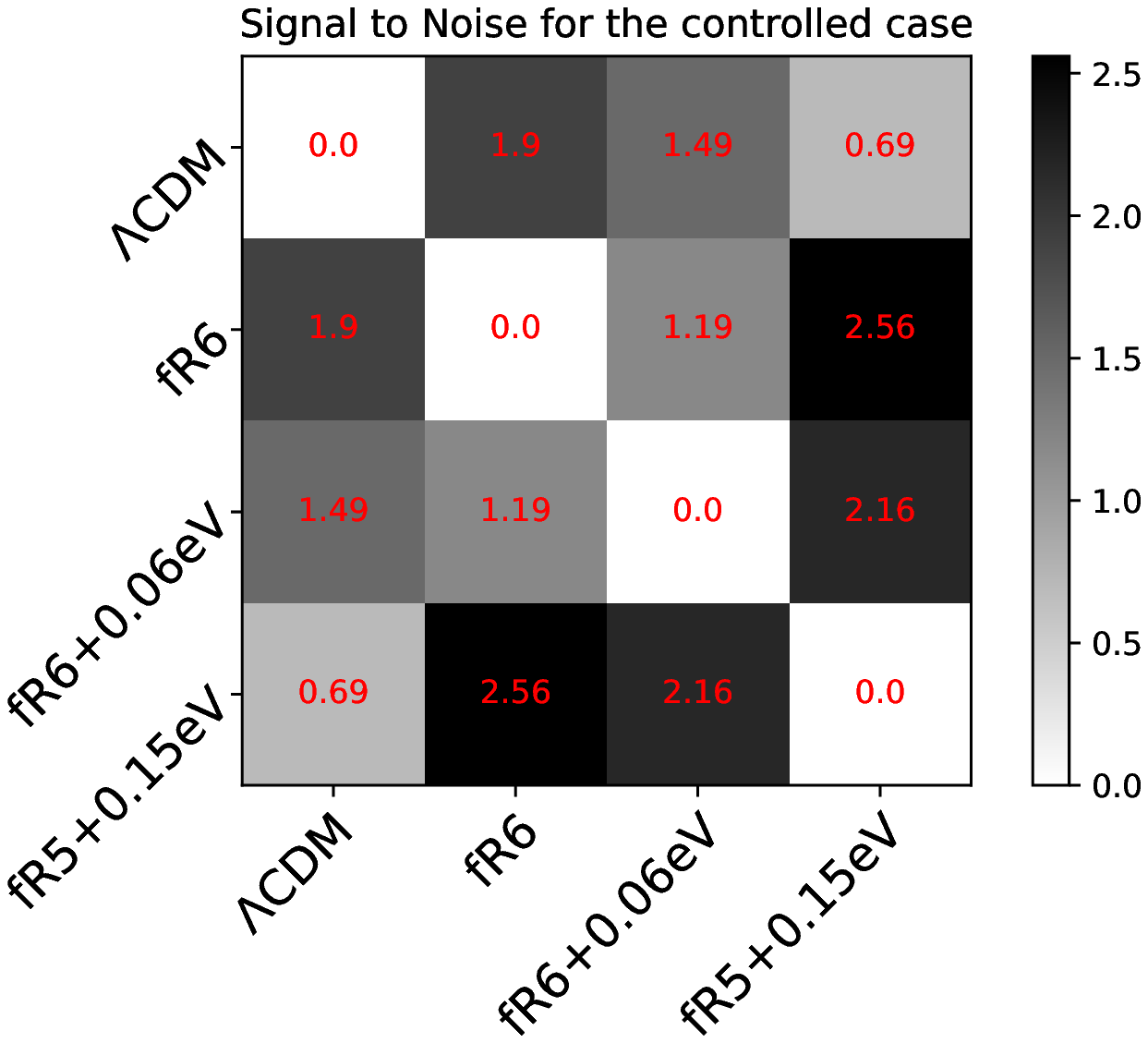}
\caption{Same as Figure \ref{fig:cov} but from the controlled samples.}
\label{fig:cov_sync}
\end{center}
\end{figure}

Figures \ref{fig:cross_sync}-\ref{fig:cov_sync} plot the same as Figures \ref{fig:cross}-\ref{fig:cov} but from the controlled subsamples.  
As can be seen, the signal to noise ratios from the controlled samples diminish down to insignificant values for all of the cases.  The highest 
value of $D/\sigma_{D}$ is found to be $2.56$ from the $\frs$ and $\frf$ pair, while the other degenerate pairs, $\lcdm$ and $\frss$ yield only $1.49$, respectively. 
Nonetheless, we suspect that these lower values of $D/\sigma_{D}$ should be ascribed to the much smaller sizes of the controlled samples compared with the original ones 
(see Table \ref{tab:sig}), noting that the trend of the variation of $\eta(r)$ and its two parameters with the four models seems to be robust, insensitive to 
whether or not the samples are original or controlled. 

\section{Summary and Conclusion}\label{sec:con}

We have put forth two diagnostics based on the shape orientations of group/cluster halos to break the dark sector degeneracy between the standard $\lcdm$ and non-standard 
$\nucdm$+$f(R)$ cosmologies.  One is the statistical tendency of the halo shape orientations being preferentially aligned with the directions of minimum matter compression in 
parallel to the minor principal directions of the linearly reconstructed tidal fields, while the other is their spatial cross-correlations.  
Analyzing data subsets from the {\small DUSTGRAIN}-{\it pathfinder} simulations performed for the $\lcdm$, $\frs$, $\frss$, and $\frf$ cosmologies, 
we have determined the first diagnostics at $z=0$.  The numerical results have been compared with the analytic single parameter formula developed by \citet{lee19} under the assumption 
that the intrinsic shape alignments of the DM halos originate from the anisotropic occurrence of merging events along the cosmic 
web \citep[e.g.,][and references therein]{wes94,fal-etal02,KE05,lib-etal13b,wit-etal19}.  

Finding an excellent agreement between the analytical and numerical results, we have shown that the best-fit parameter of the analytic formula very effectively quantifies
 the significant differences in the alignment strengths between the degenerate models. The difference in the best-fit value of the parameter has turned out to be as significant 
 as $6\sigma$ ($10\sigma$) between the $\lcdm$ and $\frss$ (between the $\frs$ and $\frf$) cosmologies, despite that the two models share almost the same growth histories and 
 same normalization amplitude of the linear density power spectrum. It has been also confirmed that this result, the success of the analytic formula and its potential 
 to discriminate the degenerate models from each other, is robust against the variation of the smoothing scale of the tidal fields from $\rf=3\dunit$ to $8\dunit$. 

Meanwhile, expecting that the large-scale coherence of the linear tidal fields would induce the shape orientations of the massive halos to be spatially cross-correlated 
\citep{wes94,atl-etal06,OT00,fal-etal02,KE05,sma-etal12},  we have devised two-parameter formula expressed in terms of the linear density two-point correlation function 
and its third and fifth moments for the halo shape-shape correlations.  
It has been found that the differences in the two best-fit parameters are as substantial as $\sim 1.49\sigma_{D}$ ($\sim 2.56\sigma_{D}$) between the $\lcdm$ and $\frss$ 
(between the $\frs$ and $\frf$) cosmologies, when no prior information on the background cosmology has been used to evaluate the analytic formula and 
the halo samples have been controlled to have identical mass and axial ratio distributions.  This result indicates a much larger sample of the group and cluster halos will be required 
to test the potential of the second diagnostics as a cosmological discriminator. 

It is worth discussing why we have utilized the massive group/cluster halos rather than the galactic counterparts as the main targets for these new diagnostics. 
In the original work of \citet{chu-etal21} who for the first time suggested the halo intrinsic shape alignments as a test of $f(R)$ gravity, their analysis was made 
exclusively on the galactic scale  under the assumption that the intrinsic shape alignments of galactic halos are observable since the shape orientations of 
the DM components of galactic halos should be well aligned with those of their observable stellar components. 
However, a recent numerical analysis based on high-resolution hydrodynamical simulations revealed that this assumption cannot be justified on the galactic scale, 
witnessing significant misalignments between the shape orientations of the DM and stellar components of the galactic halos \citep{LM22}. 
The same numerical analysis also revealed that the hot gas and DM components exhibited strong shape alignments with each other and that the alignments 
become stronger on larger mass scales. Given these numerical clues along with the fact that the shape orientations of galaxy groups and clusters are often determined 
from the distributions of their hot gas components \citep[e.g.,][]{zar-etal01},  the intrinsic shape alignments of the group/cluster halos should be a better 
indicator of the nature of gravity and DM than those of the galactic halos. 

Although the galaxy groups/clusters are much less abundant than the galaxies, which is likely to cause larger statistical errors, another merit of using them 
as the main targets is that their shape orientations reflect much better the anisotropic merging along the large-scale filamentary web \citep{atl-etal06,hah-etal07}.  The analytic 
formulae expressed in terms of the linear quantities must work better in approximating the intrinsic spin alignments and spatial correlations on the group and cluster scales 
\citep{lee19} than on the galactic scales.  Moreover, our diagnostics require only single redshift observations at $z=0$, unlike the previously suggested probes to break the dark sector degeneracy, 
like nonlinear growth rate, nonlinear redshift distortions, evolution of the drifting average coefficient of the isolated cluster abundance, high order weak lensing statistics, 
and high-redshift size distribution of cosmic voids,  all of which require multi redshift observations \citep[e.g.,][]{pee-etal18,hag-etal19b,gio-etal19,wri-etal19,ryu-etal20,con-etal21}. 

Notwithstanding, our new diagnostics bear a couple of momentous limitations as a cosmology discriminator if they are to be applied to observational data. 
First, it is still quite difficult to determine the shape orientations of the group/cluster halos with high accuracy in practice, since what is readily measurable from the optical, X-ray emission and 
Sunyaev-Zeldovich effect surveys \citep{bol-etal66,SZ69,xraybook88,YG02} is the two dimensional images of their baryonic components projected onto the plane 
of sky, which themselves suffer from the nonlinear redshift-space distortion effect \citep{jac72,kai87,ham98}.  Although the complementary method based on the strong and weak 
gravitational lensing effects \citep{bar10} are often used to directly measure the three dimensional shapes of cluster halos, it can be applied only to very massive clusters with mass 
$\gtrsim 10^{15}\munit$ that can produce lensing signals strong enough to trace the dark matter distribution inside the cluster virial radii \citep[see][for a review]{KN11}. 
For this reason, it would be highly desirable to model how the shape tracers of the group/cluster halos rather than their shapes themselves are spatially correlated and to investigate 
whether or not the tracer-tracer correlations can break the dark sector degeneracy.  This investigation, however, would require to examine first the dependence of the shape-tracer relation 
on the background cosmology. Besides, it would be much harder to find an analytic formula expressed in terms of the linear quantities 
for the tracer-tracer correlations, given that the shape tracers are expected to be nonlinearly biased \citep[][and references therein]{har-etal21}. 

The second limitation comes from the fact that our diagnostics is {\it interdependent} rather than independent on the other diagnostics. In other words, although it can play the 
{\it complementary} role of distinguishing between the non-standard cosmologies degenerate with the standard one whose initial conditions are all specified, it may not be powerful 
enough to independently constrain the key cosmological parameters for itself.  Since the difference in the halo shape correlations among the $\nucdm$+f(R) models might be 
reproduced by the changes of the six cosmological parameters within the standard paradigm, it would be idealistic to explore how our diagnostics would behave in a much 
larger parameter space for the assessment of its true potential as a cosmological probe.  Our future work will be in the direction of performing these more comprehensive work 
to overcome the limitations of our new diagnostics.

\acknowledgments

We are grateful to an anonymous referee whose constructive criticisms and many useful comments helped us significantly improve the original manuscript.  
JL acknowledges the support by Basic Science Research Program through the National Research Foundation (NRF) of Korea 
funded by the Ministry of Education (No.2019R1A2C1083855). MB acknowledges support by the project "Combining Cosmic Microwave Background 
and Large Scale Structure data: an Integrated Approach for Addressing Fundamental Questions in Cosmology", funded by the 
MIUR Progetti di Ricerca di Rilevante Interesse Nazionale (PRIN) Bando 2017 - grant 2017YJYZAH. 
MB also acknowledges the use of computational resources from the parallel computing cluster of the Open Physics Hub\\
(\url{https://site.unibo.it/openphysicshub/en}) at the Physics and Astronomy Department in Bologna.

\clearpage
\clearpage
\clearpage
\clearpage
\clearpage
\clearpage
\clearpage
\clearpage
\clearpage
\clearpage
\clearpage
\clearpage
\clearpage
\clearpage
\clearpage
\clearpage

\clearpage
\begin{deluxetable}{ccccccc}
\tablewidth{0pt}
\tablecaption{Numbers of selected group/cluster halos for each cosmology.}
\setlength{\tabcolsep}{3mm}
\tablehead{Model & $|f_{R0}|$ & $\sum m_{\nu}$ & $\sigma_{8}$ & $N_{8.1}$ & $N_{10.1}$ & $N_{\rm sync}$\\
& & $[{\rm eV}]$ & & & &}
\startdata
$\Lambda$CDM		& - 					& 0.0		& 0.847 & $217567$ & $159969$ & $208500$\\
fR6							& $10^{-6}$	& 0.0 		& 0.861 & $242418$  & $176106$ & $208500$\\
fR6+0.06eV			& $10^{-6}$	& 0.06		& 0.847 & $238003$  & $173164$ & $208500$\\
fR5+0.15eV			& $10^{-5}$	& 0.15		& 0.864 & $226161$  & $168259$ & $208500$\\
\enddata
\label{tab:sig}
\end{deluxetable}

\end{document}